\documentclass[10pt]{article}

\pdfoutput=1
\usepackage[pdftex]{color}
\usepackage{amssymb}
\usepackage{amsthm} 
\usepackage{amsmath}
\usepackage{latexsym}
\usepackage{amscd}
\usepackage{graphicx}
 \usepackage{longtable}
 \usepackage{array}
\usepackage{amssymb}
\usepackage[pdftex, colorlinks=true, citecolor=green]{hyperref}
\usepackage{lscape}
\usepackage{setspace}
\usepackage{multirow}
\usepackage{wrapfig}

\newcommand{\ds}{\displaystyle}

\newcommand{\ben}{\begin{equation}}     %equation
\newcommand{\eeqn}{\end{equation}}
\newcommand{\bey}{\begin{eqnarray}}
\newcommand{\eey}{\end{eqnarray}}

\begin{document}

\begin{flushleft}
{\Large
\textbf{Effects of cortisol time-scales in a model of emotion regulation}
}

Michael Buccino$^1$, Josh Ginart$^1$, Eva Kaslik$^2$, Anca R\v{a}dulescu$^3$

$^1$\emph{Department of Physics and Astronomy, SUNY New Paltz}

$^2$\emph{Department of Computer Science, West University of Timi\c{s}oara}

$^3$\emph{Department of Mathematics, SUNY New Paltz}

\end{flushleft}

\begin{abstract}
Stress reshapes cortico-limbic dynamics across multiple timescales, with important consequences for healthy and pathological arousal responses. We develop and analyze a Wilson-Cowan model of prefrontal-amygdala circuitry, incorporating excitatory and inhibitory populations involved in emotional regulation. The effects of stress cortisol are represented as history-dependent modulations of two stress-sensitive interactions: recurrent prefrontal excitation and inhibition of basolateral amygdala principal cells.

Using numerical simulations and bifurcation analysis, we examine how this self-interacting system organizes transitions between stable activity and infra-slow oscillations, and how this structure changes with cortisol-related temporal memory. To distinguish the effects of memory temporal distribution from those of its timescale, we compare weak and strong Gamma kernels with a discrete-delay limit, representing progressively different ways in which past activity can influence current coupling. Introducing temporal memory reshapes the parameter region supporting oscillations and alters their persistence. Weak and strong Gamma kernels produce remarkably similar global dynamics, whereas the discrete-delay limit generates a substantially richer bifurcation structure, including quasiperiodic activity.

Together, these results suggest that the dynamical effects of cortisol depend not only on the characteristic timescale of its action, but also on the shape of the temporal memory through which past neural activity influences the present. Different forms of temporal integration may provide a mechanism by which neural systems implement distinct functional dynamics, while changes in this integration could promote atypical or pathological rhythmic regimes.
\end{abstract}

\section{Introduction}

\subsection{Modeling the role of the prefrontal-limbic loop in emotion regulation}

A substantial body of empirical and theoretical work has long pointed to interactions among limbic and prefrontal regions as central to effective emotion regulation~\cite{ochsner2012functional,buhle2014reappraisal_meta,underwood2021networks}. More specifically, the ability to engage and disengage emotional responses appears to depend on how amygdala subcircuits and prefrontal control systems interact—the strength and direction of those interactions, and how they change with context. These relationships are often described in terms of functional or effective connectivity and top-down gating~\cite{banks2007amygdala_frontal,berboth2021ppi_meta,morawetz2017effective,gold2015threat_anxiety_fc}. In particular, amygdala reactivity and cortical regulatory activity tend to vary together in ways that are linked to threat appraisal, reappraisal success, anxiety, and stress coping, suggesting that these behaviors are better understood through the interaction between the two systems than through activity levels in either region alone~\cite{morawetz2017effective,morawetz2016dorsal_ventral_ec,gaffrey2021child_amygdala_fc_er,gold2015threat_anxiety_fc}. These considerations motivate dynamical systems models in which coupling parameters encode the net influence of one population or region on another, and where changes in these influences can reorganize circuit behavior across regimes associated with adaptive versus pathological processing~\cite{wilson1972excitatory,berboth2021ppi_meta}.

Stress physiology offers a well known route by which context can reshape circuit interactions. The glucocorticoid hormone cortisol is released in response to stressors, and extensive evidence indicates that cortisol influences emotional learning and regulation through actions distributed across limbic and cortical targets~\cite{mcewen2016stress,arnsten2009stress,sarabdjitsingh2012stress,van2013modulatory}. Importantly, cortisol effects are not instantaneous: they include multiple temporal components (often described as faster and slower actions, including rapid neuromodulatory-like effects and slower transcriptional/structural influences) and therefore reflect an integration of recent history rather than a single time point~\cite{joels2013interplay,joels2024emerging,van2013modulatory,strelzyk2012tune}. As a consequence, cortisol can alter effective interactions within cortical-amygdala circuits over relatively slow timescales. The consequences of these changes depend on when cortisol acts, how strongly, and for how long (acute versus sustained/chronic profiles), and may include heightened amygdala reactivity, altered recovery from stress, or weaker cortical regulation~\cite{arnsten2009stress,mcewen2016stress,pan2023difference,langer2025effects}.

Alongside stimulus-locked neural responses, emotional processing unfolds against a slowly varying internal context -- observed as very low frequency (or \emph{infra-slow}) fluctuations -- describing network engagement on longer timescales than moment-to-moment reactivity~\cite{palva2012infra,watson2018infraslow}. To capture emotionally salient cortico-limbic activity at this time scale, we used a mean-field firing-rate model of the interacting neuronal populations in the key amygdala and cortical subcircuits relevant to stress regulation~\cite{wilson1972excitatory}. The model emphasizes two ideas. First, regulation depends on effective coupling between populations, so changes in connection strengths can reorganize system-level dynamics and alter the time spent in distinct operational regimes~\cite{wilson1972excitatory}. Second, cortisol influences the circuit by acting as a history-dependent modulator of those coupling parameters rather than as a direct driver of moment-to-moment oscillations~\cite{joels2013interplay,joels2024emerging,arnsten2009stress,van2013modulatory}. To capture the temporal integration of cortisol, we represent its influence through a distributed delay, implemented as a convolution of recent circuit activity with a delay kernel that encodes physiologically motivated timescales~\cite{joels2013interplay,van2013modulatory}. This yields a compact, analytically tractable mechanism by which recent arousal can reshape the effective interaction strengths within the network.

This framework allows us to ask: (i) what patterns of activity arise from the interaction of amygdala and cortical populations as coupling strengths vary, (ii) how the model can generate repeated alternations between periods of heightened limbic arousal and stronger regulatory control, and (iii) how cortisol timing -- through distributed-delay modulation of coupling -- can shift the boundaries between these behaviors, alter the persistence and duty cycle of the oscillations, and change how long the system remains in more reactive versus more regulated phases. By explicitly separating intrinsic cortico-limbic dynamics from slower endocrine modulation, the model is designed to connect two empirically grounded facts: emotion regulation depends on coordinated prefrontal-amygdala interactions, while cortisol reshapes those interactions in a strongly time-dependent manner \cite{banks2007amygdala_frontal,morawetz2017effective,pan2023difference,joels2013interplay}.

\subsection{Oscillations and rhythms}

Before describing the specific circuit components and projections, we first discuss the timescale of the dynamics addressed by the model in the context of stress-sensitive emotion regulation. We aim to study mean-field interactions between functional neural populations, with the model variables representing general recruitment of excitatory and inhibitory populations (rather than individual neuronal activity), and the coupling weights representing their effective population-level interactions. At the population scale considered here, we focus on slow changes in these population activities and interactions, occurring over seconds to minutes, a timescale relevant to the evolution of cognitive and affective states and consistent with slow fluctuations observed in population-level neuroimaging signals. The effects of cortisol are incorporated in a history-dependent manner, so that accumulated stress exposure influences how the neural populations interact over time.

Many neural recordings show a prominent very low frequency component, often discussed as infra-slow activity/oscillations with characteristic timescales of tens of seconds to minutes; in electrophysiology this band is commonly described around 0.01-0.1 Hz and treated as a widespread feature of brain dynamics~\cite{palva2012infra}. In the cortex (including prefrontal recordings), these slow components appear in raw voltage signals and have been linked to the slow modulation of cortical excitability~\cite{vanhatalo2004infraslow}. They have also been associated with behaviorally relevant variability in neural activity and cognitive performance, with the phase of these slow fluctuations predicting changes in behavior over time~\cite{monto2008very}. In multimodal work, infra-slow EEG fluctuations covary with resting-state network dynamics in fMRI, supporting the interpretation that these slow components track variations in network engagement that organize faster processing~\cite{hiltunen2014infraslow}. 

While there are studies directly supporting infra-slow oscillations in amygdala subregions, the strongest evidence is indirectly provided by the robust involvement of the amygdala in low-frequency intrinsic coupling with medial prefrontal regions in resting-state fMRI. Individual differences in amygdala-mPFC coupling are related to affective phenotypes such as anxiety and to emotion-regulatory capacity (see for example ~\cite{kim2011anxiety}). Direct invasive evidence for infra-slow dynamics in the amygdala also exists, although it is more limited and method-dependent than in cortex. For example, BLA field recordings in slice have revealed spontaneous periodic events occurring at frequencies below 0.1 Hz~\cite{bratsch2024synaptic}. In humans, depth recordings have found robust 0.01-0.1 Hz infra-slow activity in stereoelectroencephalography during temporal lobe seizures, although the anatomical specificity varies by implantation coverage~\cite{thompson2016ictal}.

From a modeling standpoint, the relevance of infra-slow oscillations is that they appear empirically as slow population-state fluctuations, evolving over tens of seconds to minutes and associated with changes in excitability, engagement, and intrinsic network organization~\cite{palva2012infra}. In this context, the Wilson-Cowan framework is a natural fit because its state variables already operate at the appropriate biological level: they represent the aggregate recruitment of interacting excitatory and inhibitory populations, rather than the activity of individual neurons~\cite{wilson1972excitatory}. The coupling terms can then be interpreted as effective population-level feedback gains, describing in general how activity in one region influences recruitment in another: how strongly local excitation reinforces itself, how strongly inhibition constrains activity, how efficiently excitation recruits inhibition, and, in multi-region extensions, how activity in one area biases the recruitment of excitatory or inhibitory populations in another. This population-level description is particularly relevant for infra-slow dynamics because the quantities represented by these effective gains and operating states are precisely those thought to vary slowly in the underlying physiology. Infra-slow components have been associated with slow modulation of cortical excitability and responsiveness, and their phase can organize the amplitude of ongoing neural activity~\cite{vanhatalo2004infraslow}. A Wilson-Cowan model therefore provides a natural framework for studying how slow changes in recruitment, responsiveness, and inter-population feedback can shape the evolution of neural population states over tens of seconds to minutes.

Wilson-Cowan type population dynamics have been explicitly paired in existing modeling work with fMRI, MEG, EEG, and invasive field recordings (which are the modalities in which infra-slow activity has been observed). On the fMRI side, connectome-scale simulations have modeled resting activity using networks of noise-driven Wilson-Cowan oscillators to reproduce large-scale resting-state structure~\cite{deco2009key}, and Wilson-Cowan neuronal equations have been inserted directly into the fMRI effective-connectivity framework~\cite{sadeghi2020dynamic}. Related whole-brain fMRI modeling work also uses large-scale Wilson-Cowan neural mass networks with hemodynamic forward models to generate BOLD-like signals and compare connectivity structure across conditions~\cite{masharipov2024comparison}. On the MEG side, whole-brain cortical networks have been simulated using the Wilson–Cowan neural mass model and then evaluated against empirical MEG functional connectivity measures, including envelope-based metrics~\cite{abeysuriya2018biophysical}. On the electrophysiology side, Wilson-Cowan/Amari style neural field formulations have been used as parametric generative models for electrophysiological data, with parameters estimated from data in a data-assimilation framework~\cite{freestone2011data}. Closely related neural field models have been used within Dynamic Causal Modeling to estimate model parameters from EEG or LFP spectra~\cite{pinotsis2012dynamic}. Some broader reviews summarize how neural field models are paired with invasive (ECoG/LFP) and non-invasive (EEG/MEG) data in this Bayesian inversion setting~\cite{moran2013neural}.

Finally, the tie between infra-slow rhythms and emotion regulation has been identified in a variety of studies, and the evidence lines up naturally with our modeling choices. Infra-slow activity behaves like a meaningful (rather than incidental) internal state variable in cortico-limbic circuits~\cite{palva2012infra,watson2018infraslow}. In humans, intrinsic 0.1Hz dynamics show emotion-relevant structure: for example, amygdala-prefrontal phase coupling of intrinsic 0.1Hz BOLD oscillations is stronger in individuals with elevated state anxiety, directly linking infra-slow rhythms to this affective phenotype~\cite{pfurtscheller2018synchronization}. Infra-slow activity is relevant for regulation because it appears to reflect slow fluctuations in the brain’s baseline responsiveness~\cite{palva2012infra,watson2018infraslow}. When the infra-slow component is in one phase, the same circuitry may be more readily recruited and thus more prone to reactive responses; in the opposite phase, recruitment may be reduced, favoring more stable, controlled processing~\cite{vanputten2015infraslow,de2019infraslow}. Intracranial recordings support this interpretation by showing that infra-slow rhythms are accompanied by slow, quasi-periodic modulation of cortical excitability, suggesting that neural processing occurs under a slowly shifting operating point over tens of seconds to minutes~\cite{vanhatalo2004infraslow,vanputten2015infraslow}. Consistent with this view, infra-slow activity in the 0.01-0.1 Hz range predicts both slow fluctuations in behavioral performance (on one hand) and the expression of faster activity (on the other hand), supporting a functional role for these slow states rather than their interpretation as noise or as an incidental drift~\cite{hultman2016dysregulation,monto2008very,thompson2014phase,sihn2022differential}. 

In animal work, the link between slow dynamics and emotion is explicit at the circuit level. Hultman et al. describe stress-induced emotional pathology in terms of disrupted slow-evolving network dynamics across cortico-limbic circuits, and show that prefrontal circuitry regulates these patterns in ways that suggest maladaptive emotional behavior~\cite{hultman2016dysregulation}. It is also suggested that these slow rhythms are shaped by coupling: connectivity helps determine whether infra-slow activity synchronizes, propagates, or fragments across a network~\cite{hiltunen2014infraslow,grooms2017dc}. Consistent with this, amygdala-prefrontal coupling near 0.1 Hz has been shown to strengthen with anxiety~\cite{pfurtscheller2018synchronization}, while changes in prefrontal circuit function can reorganize slow-evolving cortico-limbic dynamics rather than simply altering local activity~\cite{hultman2016dysregulation}. Together, these findings support the modeling view that infra-slow activity can provide a substrate for endogenous state cycling in cortico-limbic circuits, that feedback coupling can reorganize this cycling by altering coordination across regions, and that these dynamics can be meaningfully related to emotional regulation through changes in intrinsic coupling and network state occupancy~\cite{palva2012infra,hiltunen2014infraslow}.

\subsection{Impact of stress cortisol}

The primary aim of our project is to understand the delayed, cumulative effects of stress cortisol on prefrontal-limbic regulation. Stress activates the hypothalamic-pituitary-adrenal (HPA) axis and produces an elevation in circulating cortisol, the neural consequences of which unfold over multiple time scales. Cortisol action is not determined only by the magnitude of the hormonal response, but also by its timing. Rapid, largely non-genomic corticosteroid effects can emerge within minutes, while slower receptor-mediated genomic and downstream cellular effects develop over tens of minutes to hours and may persist after circulating cortisol has begun to decline \cite{joels2013interplay,joels2024emerging}. These temporal components are not simply weaker or stronger versions of the same effect. Experimental studies increasingly suggest that cortisol can influence the same cognitive and cortico-limbic processes differently depending on when they are measured relative to the stress response.

This timing dependence is particularly clear in studies of emotion regulation. Pan et al. compared cortisol effects approximately 30 and 90 minutes after administration and found qualitatively different patterns across the two windows \cite{pan2023difference}: rapid cortisol action increased amygdala and dorsolateral prefrontal activation and their functional connectivity during emotion regulation, whereas slower cortisol effects reduced these responses and were associated with improved down-regulation of negative emotion. Similarly, cortisol administered 30 minutes before emotional-memory encoding impaired contextualization, whereas administration 210 minutes before encoding enhanced it \cite{vanAst2013Timing}. Importantly, however, observations of heightened versus reduced effects at selected time points provide only snapshots of an underlying temporal process, which may be substantially more complex than a simple transition from one functional state to another. Together with a broader literature on rapid and slow corticosteroid signaling, these findings indicate that the functional impact of cortisol depends strongly on the temporal relationship between stress exposure, hormonal action, and the neural process being measured \cite{joels2013interplay,joels2024emerging}.

At the level of the circuitry modeled here, this temporal organization is also reflected in distinct stress-sensitive mechanisms. In the basolateral amygdala, rapid glucocorticoid actions can suppress GABAergic inhibition of principal neurons through endocannabinoid-mediated signaling, producing a relatively fast disinhibitory effect \cite{di2016acute,atsak2015endocannabinoid,joels2024emerging}. Prefrontal effects are expressed over longer time scales and, under sustained or repeated stress, include changes in pyramidal-cell structure and synaptic organization that can weaken recurrent prefrontal interactions and regulatory function \cite{arnsten2009stress,liston2006stress,mcewen2013brain,mcewen2016stress,radley2006repeated}. We will therefore need to distinguish between a faster cortisol-sensitive modulation of amygdala inhibition and a slower modulation of recurrent prefrontal excitation. We represent these
pathways by separate fast and slow filtered signals,
defined below.

The temporal dependence of these effects motivates treating cortisol action as a memory of recent activity rather than as an instantaneous input. In our model, the current stress-related modulation depends on a weighted history of previous amygdala activity, so that both \emph{how far into the past} the relevant activity lies and \emph{how that past activity is distributed in time} can influence the present coupling strengths. The delay kernel therefore represents the temporal profile over which previous arousal contributes to the current cortisol-mediated effect. By varying the form of this temporal weighting, we can ask whether cortico-limbic dynamics depend only on the overall timing of cortisol action, or also on whether its influence is broadly distributed over recent history or concentrated around a more specific previous time.

\section{Modeling methods}
\subsection{The circuit: regions and projections}

With our model, the aim is to define a minimal cortico-limbic context in which the sign and strength of key projections capture established motifs of reactivity, gating, and regulation of emotion. The four-region architecture we are considering provides a concrete and simple circuit structure, on which we will build our dynamical analysis and test our hypotheses. The four interacting variables are as follows:\\

\noindent \emph{\textbf{Pyramidal cortical neurons ($P$).}} Excitatory glutamatergic neurons in the prefrontal cortex form recurrent microcircuits and provide the main cortical output. They integrate local and long-range inputs and project to subcortical structures, including the amygdala~\cite{douglas2004neuronal,harris2015neocortical}.

\vspace{2mm}
\noindent \emph{\textbf{Somatostatin-expressing cortical interneurons ($S$).}}
GABAergic interneurons expressing somatostatin preferentially target distal dendrites of pyramidal neurons, shaping dendritic integration and providing both feedback and feed-forward inhibition~\cite{urban2016somatostatin,tremblay2016gabaergic}.

\vspace{2mm}
\noindent \emph{\textbf{Cortico-amygdala principal neurons ($C$).}} Glutamatergic principal neurons in the basolateral amygdala (BLA) receive cortical input and send reciprocal excitatory projections back to prefrontal cortex, forming the major excitatory pathway for PFC-amygdala communication~\cite{likhtik2015amygdala,little2012subcellular}.

\vspace{2mm}
\noindent \emph{\textbf{Basolateral amygdala interneurons ($B$).}} Local GABAergic interneurons, including PV+ and SOM+ subtypes, are strongly recruited by BLA principal neurons and provide potent inhibition onto them, regulating excitability, synchrony, and gating of amygdala output~\cite{spampanato2011interneurons,wolff2014amygdala}.\\

\noindent Below, we describe the main projections within this circuit, organized by their target:\\

\noindent \emph{\textbf{Inputs to cortical pyramidal neurons ($P$).}} Cortical pyramidal cells receive strong recurrent excitatory inputs from other pyramids, forming the canonical glutamatergic microcircuit of the prefrontal cortex. This recurrent excitation is thought to underlie sustained activity and amplification of cortical signals, supporting functions such as working memory and top-down control~\cite{douglas2004neuronal,harris2015neocortical}. They are also regulated by inhibitory inputs from local somatostatin-expressing (SST) interneurons, which target pyramidal dendrites. By controlling dendritic integration, SST interneurons gate the influence of long-range inputs and shape the excitability of pyramidal cells~\cite{urban2016somatostatin,tremblay2016gabaergic}. Finally, pyramidal neurons in the prefrontal cortex receive excitatory projections from basolateral amygdala (BLA) principal neurons. These glutamatergic afferents can directly depolarize pyramidal cells, but their net effect is often shaped by concurrent recruitment of local inhibitory interneurons that provide feed-forward inhibition. Hence the BLA$\rightarrow$PFC pathway conveys excitatory drive while also modulating pyramidal output through inhibitory control~\cite{little2012subcellular,mcgarry2016inhibitory,likhtik2015amygdala}.

%%%

\vspace{2mm}
\noindent \emph{\textbf{Inputs to basolateral amygdala principal neurons (C).}}
Principal glutamatergic neurons in the basolateral amygdala (BLA) receive excitatory projections from pyramidal neurons in the prefrontal cortex. These cortico-amygdala inputs provide a major top-down signal, allowing cortical activity to directly shape the amygdala's excitability and plasticity during emotional processing \cite{mcdonald1998cortical,likhtik2015amygdala}. Within the BLA itself, principal cells also receive strong recurrent excitatory input from other principal neurons, forming a local excitatory network that supports amplification of incoming signals and contributes to associative learning \cite{sah2003amygdaloid,pape2010plastic}. Counterbalancing these excitatory drives, BLA principal neurons are tightly controlled by local inhibitory interneurons. Both parvalbumin- and somatostatin-expressing interneurons are robustly recruited by principal cell activity and provide potent inhibition back onto them. This recurrent inhibitory loop regulates firing patterns, synchrony, and gating of information flow from the amygdala to the cortex \cite{spampanato2011interneurons,wolff2014amygdala}.

%%%

\vspace{2mm}
\noindent \emph{\textbf{Inputs to cortical SST interneurons (S).}} Somatostatin-expressing interneurons in the prefrontal cortex receive strong excitatory inputs from local pyramidal neurons, which recruit them as part of feedback inhibitory loops. In this way, pyramidal activity engages SST interneurons to regulate dendritic integration and prevent runaway excitation in cortical microcircuits~\cite{tremblay2016gabaergic,urban2016somatostatin}. In addition to local recruitment, projections from basolateral amygdala (BLA) principal neurons also synapse onto cortical SST interneurons. This pathway provides a form of feed-forward inhibition: excitatory signals from the amygdala preferentially activate inhibitory interneurons in prefrontal cortex, which in turn suppress pyramidal excitability. Such organization allows BLA input to modulate cortical processing not only by direct excitation of pyramids but also by gating through inhibitory circuits~\cite{little2013synaptic,likhtik2015amygdala}. Finally, SST interneurons exhibit a degree of self- and network-mediated inhibition, including inhibitory interactions with other interneuron classes such as parvalbumin (PV) cells. While not explicitly represented in the reduced four-node model, these interactions contribute to shaping the temporal dynamics and gain of SST activity~\cite{markram2004interneurons,tremblay2016gabaergic}.
%%%

\vspace{2mm}
\noindent \emph{\textbf{Inputs to BLA interneurons (B).}} Local interneurons in the basolateral amygdala are primarily recruited by excitatory input from BLA principal neurons. This strong local drive ensures that any increase in principal cell activity is rapidly counterbalanced by inhibition, forming a tightly coupled excitatory-inhibitory loop that stabilizes network dynamics and shapes the amygdala's output~\cite{spampanato2011interneurons,wolff2014amygdala}. In addition, pyramidal neurons in the prefrontal cortex provide direct excitatory projections to BLA interneurons. This pathway constitutes a form of cortico-amygdala feed-forward inhibition: cortical activity can engage inhibitory cells in the BLA, which then suppress principal neuron excitability. Such a mechanism enables the cortex to down-regulate amygdala output indirectly, providing a substrate for top-down control of emotional responses~\cite{mcdonald1998cortical,likhtik2015amygdala}. Finally, BLA interneurons also receive inputs from other local interneurons, including both PV- and SST-expressing subtypes, creating a microcircuit of inhibitory interactions that further modulates their responsiveness. While these interneuron-interneuron connections are not explicitly represented in the simplified four-node model, they contribute to shaping the temporal precision and gain of inhibition in the amygdala~\cite{spampanato2011interneurons,wolff2014amygdala}.

\subsection{The basic model and parameters}

The original Wilson-Cowan system consists of two coupled integro-differential equations, describing activity in two interacting excitatory (E) and inhibitory (I) neural populations (see [1]):
\begin{equation*}
\left\{\begin{array}{l}
\ds E(t+\tau_1)=\left(1-\int_{t-r}^tE(s)ds\right)\cdot\mathcal{S}_e\left[\int_{-\infty}^t h(t-s)\left(c_1E(s)-c_2I(s)+P_e(s)\right)ds\right]\\
\ds I(t+\tau_2)=\left(1-\int_{t-r'}^tI(s)ds\right)\cdot\mathcal{S}_i\left[\int_{-\infty}^t h(t-s)\left(c_3E(s)-c_4I(s)+P_i(s)\right)ds\right]
\end{array}\right.
\label{unsimplified}
\end{equation*}

\noindent where $r$ and $r'$ are the absolute refractory periods, $c_i>0$ are connectivity weights, and $P_e,P_i$ denote external inputs. It is additionally assumed that a cell's weighted summation of inputs is modulated longitudinally by a decreasing kernel $h(t)$, and that effective inputs are integrated through sigmoid functions $\mathcal{S}_{\beta_E,\theta_E}$, $\mathcal{S}_{\beta_I,\theta_I}$, where: 
$${\cal S}_{\beta,\theta}(x) = \frac{1}{1+\exp(-\beta(x-\theta))}-\frac{1}{1+\exp(\beta \theta)}$$

\vspace{3mm}
\noindent Via averaging and coarse-graining, the Wilson-Cowan system is best known in its simplified ODE form:
\begin{equation*}
\left\{\begin{array}{l}
\ds \dot{E}(t)=-E(t)+\left(1-rE(t) \right)\cdot\mathcal{S}_e\left(c_1E(s)-c_2I(t)+P_e(t)\right)\\
\ds \dot{I}(t)=-I(t) + (1-rI(t))\cdot\mathcal{S}_i\left(c_3E(s)-c_4I(s)+P_i(s)\right)
\end{array}\right.
\label{unsimplified}
\end{equation*}

\noindent In prior research [2], we used Wilson-Cowan coupled units to explore two different wiring schemes proposed in the empirical literature of stress regulation. We investigated the effects of uniform delays in the E/I pathways. In the current project, we refine the delays to align with the effects of slow circulating cortisol, and its effects on neural function and synaptic transmission.

\begin{figure}[h!]
\begin{center}
\includegraphics[width=0.4\linewidth]{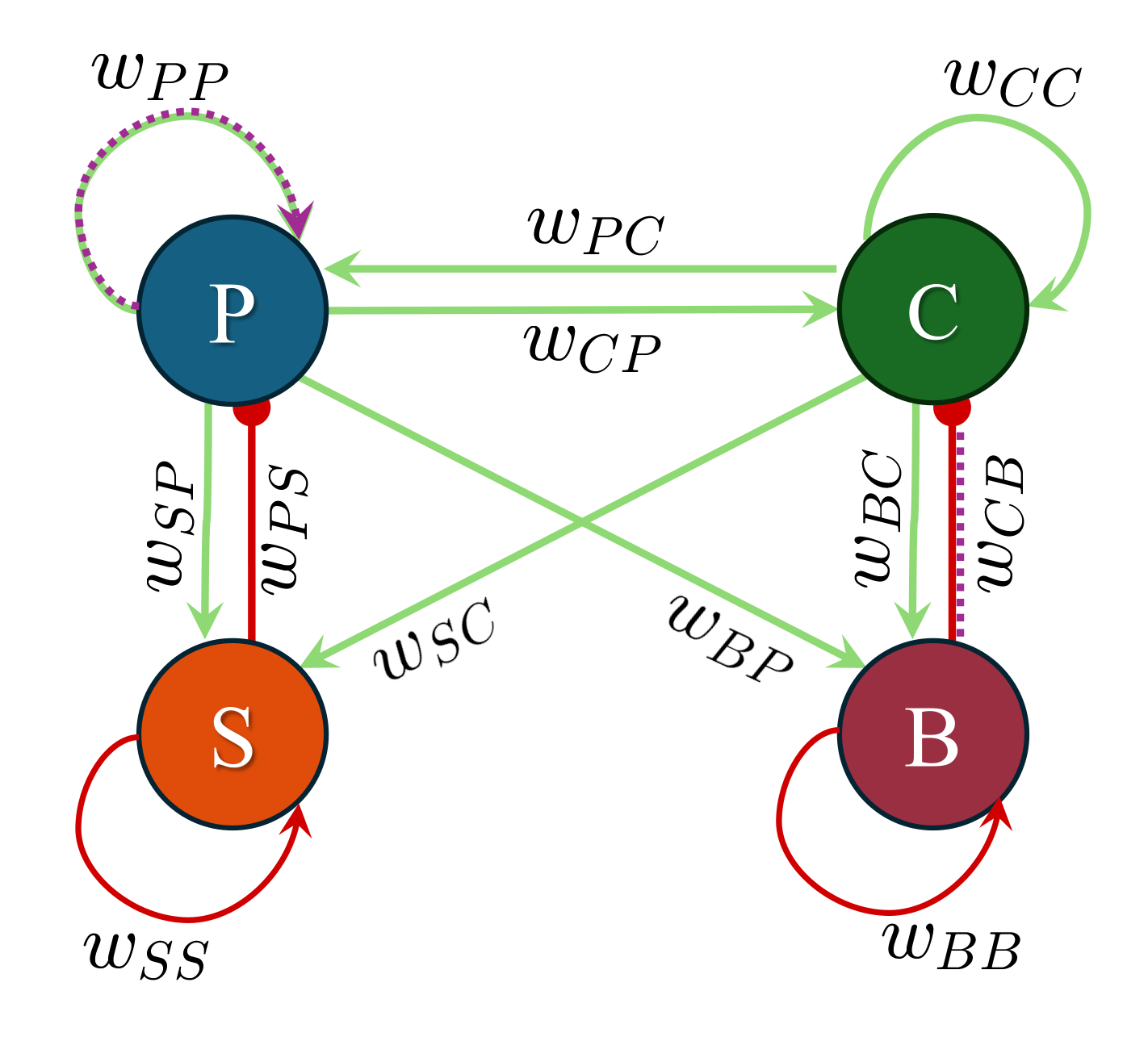}
\end{center}
\vspace{-3mm}
\caption{\small \emph{{\bf Diagram of the cortico-limbic circuit addressed by our model.} The variables $P$, $C$, $S$, and $B$ represent the mean-field activation levels in neural populations from the four regions described above, while the coefficients $w_{ij}$ represent the effective coupling strength from population $j$ to population $i$, incorporating the net population-level influence of the corresponding projection. Excitatory projections from $j$ to $i$ (with corresponding $w_{ij}>0$) are shown as green arrows, and inhibitory projections  (with $w_{ij}<0$) are shown in red. The weights $w_{PP}$ and $w_{CB}$ are singled out in particular, as the potential targets for cortisol impact.}}
\label{fig:placeholder}
\end{figure}

\noindent The basic model can be written as a system of differential equations, in which the activation variables $P$, $C$, $S$ and $B$ take values in $[-1,1]:$

%{\color{red}Eva: I see a small issue here: if we are taking the refractory factors $(1-P)$, $(1-C)$, etc., doesn't this suggest (imply) that $P,C,S,B$ should be fractions, so confined to $[0,1]$? And consequently, does it make sense to consider shifted sigmoids?}

\begin{eqnarray*}
\tau_P  \dot P(t)
&= -P(t) + \bigl(1-P(t)\bigr)\,\mathcal{S}_{\beta_P,\theta_P} \Big(w_{PP}P(t) + w_{PC}C(t) + w_{PS}S(t) + I_P \Big)\\
\tau_C \dot C(t)
&= -C(t) + \bigl(1-C(t)\bigr)\mathcal{S}_{\beta_C,\theta_C}\Big(w_{CP}P(t) + w_{CC}C(t) + w_{CB}B(t) + I_C \Big)\\
\tau_S\dot S(t)
&= -S(t) + \bigl(1-S(t)\bigr)\mathcal{S}_{\beta_S,\theta_S}\Big({w_{SP}P(t) + w_{SC}C(t) + w_{SS}S(t) + I_S}\Big)\\
\tau_B\dot B(t)
&= -B(t) + \bigl(1-B(t)\bigr)\mathcal{S}_{\beta_B,\theta_B}\Big({w_{BP}P(t) + w_{BC}C(t) + w_{BB}B(t) + I_B}\Big)
\end{eqnarray*}

More precisely,  the variables $P$, $C$, $S$ and $B$ represent signed, normalized population-recruitment deviations. Negative
values indicate recruitment below the chosen reference level and are not literal negative firing rates. This
signed interpretation is consistent with the shifted activation function used here. The baseline values and ranges of the response and coupling parameters are given in Table~\ref{param_table}. 

The stress-related modulation is represented by distributed delays driven by the activity of the amygdala principal-cell population. To distinguish the fast and slow cortisol-mediated effects introduced below, we define two filtered activity variables,
\begin{equation}
R_f(t)=\int_0^\infty g_f(\theta;n_f,\tau_f)\,C(t-\theta)\,d\theta,
\qquad
R_s(t)=\int_0^\infty g_s(\theta;n_s,\tau_s)\,C(t-\theta)\,d\theta.
\label{eq:stress-signals}
\end{equation}
The kernels $g_f$ and $g_s$ are chosen to be Gamma distributed-delay kernels of the form
\begin{equation}
g_j(\theta;n_j,\tau_j)
=
\frac{\theta^{n_j-1}e^{-\theta/\tau_j}}{\tau_j^{\,n_j}\Gamma(n_j)},
\qquad \theta\ge 0,
\qquad j\in\{f,s\}.
\label{eq:gamma}
\end{equation}
Here $n_j\in\mathbb N$ is a shape parameter and $\tau_j>0$ is a scale parameter. In particular, the mean delay is {\color{red}$\mu_j=n_j\tau_j$}.

When $n_f$ and $n_s$ are positive integers, the distributed-delay formulation can be rewritten exactly as an equivalent finite-dimensional ODE system by means of the linear chain trick. Introducing auxiliary variables $z_1,\dots,z_{n_f}$ and $y_1,\dots,y_{n_s}$, we obtain
\begin{align}
\dot z_1(t) &= \frac{C(t)-z_1(t)}{\tau_f}, \nonumber\\
\dot z_k(t) &= \frac{z_{k-1}(t)-z_k(t)}{\tau_f}, \qquad k=2,\dots,n_f,
\label{eq:chain-fast}
\end{align}
and
\begin{align}
\dot y_1(t) &= \frac{C(t)-y_1(t)}{\tau_s}, \nonumber\\
\dot y_k(t) &= \frac{y_{k-1}(t)-y_k(t)}{\tau_s}, \qquad k=2,\dots,n_s.
\label{eq:chain-slow}
\end{align}
The delayed signals are then recovered as
\begin{equation}
R_f(t)=z_{n_f}(t),
\qquad
R_s(t)=y_{n_s}(t).
\label{eq:stress-chain-output}
\end{equation}

\noindent This reformulation is especially convenient for numerical continuation and local stability analysis, since it replaces the convolution terms by an augmented ODE system. In particular, near an equilibrium, the corresponding linearization yields a finite-dimensional Jacobian and hence,  a polynomial characteristic equation. The filtered signals $R_f(t)$ and $R_s(t)$ provide biologically plausible proxies for fast and slow cortisol-mediated modulation. We model two main effects:\\

\noindent \emph{\textbf{Fast cortisol action on the BLA inhibitory pathway.}} Empirically, glucocorticoids can act within minutes in the basolateral amygdala to reduce synaptic inhibition: acute stress or glucocorticoid exposure rapidly mobilizes retrograde endocannabinoid signaling in the BLA, engaging CB1 receptors to suppress inhibitory synaptic transmission onto principal neurons, with behavioral consequences consistent with increased anxiety and facilitated fear-related processing~\cite{di2016acute,tasker2011mechanisms,joels2024emerging}. This endocannabinoid route is now widely treated as a canonical mechanism for rapid (non-genomic) corticosteroid modulation of synaptic function in stress circuits, including the amygdala~\cite{hill2012endocannabinoid,joels2024emerging,atsak2015endocannabinoid}. In our model, this corresponds to an increase in $R_f(t)$
reducing the magnitude of inhibitory $w_{CB}$, capturing stress-induced disinhibition
of the principal neurons of BLA:
%In the model, this is implemented as a \emph{time-varying} inhibitory weight:
$$w_{CB}(t) \;=\; w_{CB}\,\bigl( 1 - \beta_{CB}\, R_f(t) \bigr)$$

\noindent Here, $\beta_{CB}$ represents the strength of the fast cortisol impact, which is considered low enough to ensure that $\beta_{CB}R<1$ at all times, and $w_{CB}$ remains negative. \\

\noindent \emph{\textbf{Slow cortisol actions on prefrontal recurrent excitation.}} Empirically, chronic stress is consistently associated with reduced dendritic complexity and synaptic connectivity in prefrontal pyramidal populations, a form of structural remodeling that is widely interpreted as a slow (genomic/structural) consequence of prolonged glucocorticoid exposure and stress signaling in PFC~\cite{arnsten2009stress,liston2006stress,shansky2009stress,mcewen2013brain}. Across chronic stress paradigms in rodents, layer II/III medial prefrontal pyramidal neurons show retraction of apical dendritic arbors and related reductions in spine density, and these morphological changes track impairments in prefrontal-dependent cognitive control (e.g., set-shifting), linking circuit-level remodeling to functional loss of flexible regulation~\cite{liston2006stress,radley2008repeated}. Reviews synthesizing this literature emphasize that such PFC remodeling contrasts with stress-associated hypertrophy in some limbic structures, reinforcing the idea that chronic stress can weaken recurrent prefrontal network integrity and thereby compromise top-down control over behavior and emotion over longer timescales~\cite{arnsten2009stress,mcewen2013brain}. In our model, this effect is modeled by allowing the
excitatory recurrent coupling in $P$ to decay with the delayed cortisol
signal:
$$w_{PP}(t) \;=\; w_{PP}\,\bigl( 1 - \beta_{PP}\, R_s(t) \bigr)$$

\noindent Here, $\beta_{PP}$ represents the impact of slow cortisol, which is considered low enough to ensure that $\beta_{PP}R<1$ at all times, and $w_{PP}$ remains positive.

To compare and contextualize these two cortisol actions across various stress distribution
shapes, we considered both weak and strong Gamma kernels, as well as discretely distributed delays. In all cases, we fixed the average delay as $n_f\tau_f\in [2,8]$ (min), for the fast cortisol action, and $n_s\tau_s\in[15,120]$ min, to reflect the slower cortisol action.

\section{Numerical simulations}

For our analysis, we considered $w_{PP}$ and $w_{CB}$, the sites of cortisol action, as our key parameters. We investigate how the shape and timing of the cortisol response to stress affect the dependence of the system on these coupling strengths. The other parameters were fixed to the baseline values in Table~\ref{param_table}, and were chosen based on existing empirical and modeling ranges and ratios.\\

\noindent In setting the coupling ranges in Table~1, we follow empirical constraints on relative circuit dominance rather than treating $w_{ij}$ as literal unitary synaptic strengths. First, local recurrent excitation and local feedback inhibition are modeled as the largest-magnitude influences, consistent with extensive evidence that cortical and amygdala interactions are organized around strong local E-I loops in which inhibition rapidly tracks and counterbalances excitation, shaping gain and preventing excessive activation~\cite{isaacson2011inhibition,markram2004interneurons,xue2014equalizing}. In particular, we used broad ranges for local recurrent pyramidal recruitment $w_{PP}$ and for inhibitory control of the amygdala excitatory population through $w_{CB}$, reflecting the potency of recurrent excitation and interneuron-mediated gating in cortical and BLA microcircuits~\cite{pfeffer2013inhibition,spampanato2011interneurons,wolff2014amygdala}. Second, long-range cortico–amygdala couplings are kept nonzero but smaller in magnitude than dominant local loop terms, consistent with anatomical and physiological evidence that PFC $\leftrightarrow$ BLA projections strongly modulate excitability and learning, but operate through recruitment and gating of local microcircuits rather than replacing them as the primary stabilizing feedback~\cite{mcdonald1998cortical,little2013synaptic,likhtik2015amygdala}. Finally, the ratio choices for the stress-sensitive pathways are anchored in empirical findings that acute stress/glucocorticoids can rapidly suppress GABAergic inhibition in BLA via endocannabinoid mechanisms (motivating a sizeable dynamic range for $w_{CB}$), whereas chronic stress reliably impairs PFC recurrent integrity via dendritic retraction/spine loss and reduced efficacy of recurrent excitation (motivating a substantial but slower-varying range for $w_{PP}$)~\cite{di2016acute,arnsten2009stress,radley2006repeated,liston2006stress}.

Tonic inputs are applied only to the two excitatory principal populations ($I_P$ and $I_C$) because these cell classes receive the dominant long-range excitatory and modulatory drive and thus set the main operating point of each region, whereas interneuron populations are typically recruited in a state-dependent manner by local principal-cell activity. In the cortex, pyramidal neurons integrate corticocortical, thalamocortical, and neuromodulatory influences, while SST interneurons are strongly recruited by local pyramidal activity and participate in feedback inhibition, so it is natural to treat their baseline activation as arising primarily from $P \to S$ coupling rather than from an independent tonic input ~\cite{urban2016somatostatin,tremblay2016gabaergic,pfeffer2013inhibition}. Similarly, in the BLA, principal neurons receive substantial excitatory afferents, while local interneurons are strongly driven by principal-cell firing and provide rapid feedback inhibition; representing baseline drive at the level of $C$ and letting inhibitory recruitment arise mainly through $C \to B$ captures this organization and its disinhibitory control loops~\cite{spampanato2011interneurons,wolff2014amygdala,mcdonald1998cortical}. With this choice, $I_P$ and $I_C$ encode the net background/external tone entering each region, while interneuron activity remains a recruited control signal that scales with local excitation.

The Wilson-Cowan parameters in Table~\ref{param_table} are chosen to match well-established empirical regularities at the population level. The time constants in Table~\ref{param_table} were chosen to reflect differences in the characteristic response times of the modeled populations. In particular, the amygdala inhibitory population was assigned the fastest timescale ($\tau_B=4$), consistent with rapid local inhibitory tracking of excitatory activity and the fast kinetics of GABA$_A$-mediated inhibition. The remaining populations were assigned somewhat slower timescales ($\tau_P=8,\tau_C=10,\tau_S=12$), reflecting their comparatively slower population-level recruitment and integration~\cite{isaacson2011inhibition,markram2004interneurons,tremblay2016gabaergic}. The sigmoid parameters $\theta_i$ and $\beta_i$ are used in the original Wilson-Cowan sense as a coarse-grained description of heterogeneous threshold recruitment in a neuronal population: $\theta_i$ sets an effective operating threshold and $\beta_i$ sets a moderate gain (steep enough for meaningful modulation but not steep enough to approach a hard switch)~\cite{wilson1972excitatory,chow2020before,byrne2020next}. In neural-mass practice, such moderate slopes are also interpreted as reflecting the variance of individual firing thresholds within the population, motivating values in a mid-range (here $\beta\approx 3.5\!-\!4$) that preserve sensitivity and avoid immediate saturation across states~\cite{coronel2021cholinergic}. Setting a common $\theta=0.4$ across nodes is also a standard normalization choice that places the populations in a comparable responsive range, with differences in recruitment determined by the connectivity and gain parameters~\cite{wilson1972excitatory,byrne2020next,moran2013neural}.

\begin{table}[h!]
\begin{center}
\noindent \begin{tabular}{|l|l|l|l|l|l|l|l|}
\hline
Name & Value & Name & Value & Name & Value & Name & Value\\
\hline
$w_{PP}$ & 0-8 & $w_{PC}$ & 0.3 & $w_{PS}$ & -3 & $I_P$ & 0.18\\ 
$w_{CP}$ & 1.5 & $w_{CC}$ & 2 & $w_{CB}$ & -7-0 & $I_C$ & 0.2 \\ 
$w_{SP}$ & 2.2 & $w_{SC}$ & 1.2 & $w_{SS}$ & -0.4 & $\beta_{PP}$ & 2\\ 
$w_{BP}$ & 0.8 & $w_{BC}$ & 2.2 & $w_{BB}$ & -0.4 & $\beta_{CB}$ & 2\\  
\hline
\end{tabular}

\begin{tabular}{|l|l|l|l|l|l|}
\hline
Name & Value & Name & Value & Name & Value\\
\hline
$\tau_P$ & 8 & $\beta_P$ & 4 & $\theta_P$ & 0.4\\ 
$\tau_C$ & 10 & $\beta_C$ & 3.5 & $\theta_C$ & 0.4\\ 
$\tau_S$ & 12 & $\beta_S$ & 4 & $\theta_S$ & 0.4\\  
$\tau_B$ & 4 & $\beta_B$ & 4 & $\theta_B$ & 0.4\\ 
\hline
\end{tabular}
\end{center}
\vspace{-3mm}
\caption{\emph{\bf Model parameters.}}
\label{param_table}
\end{table}

\begin{figure}[h!]
\begin{center}
\includegraphics[width=0.8\linewidth]{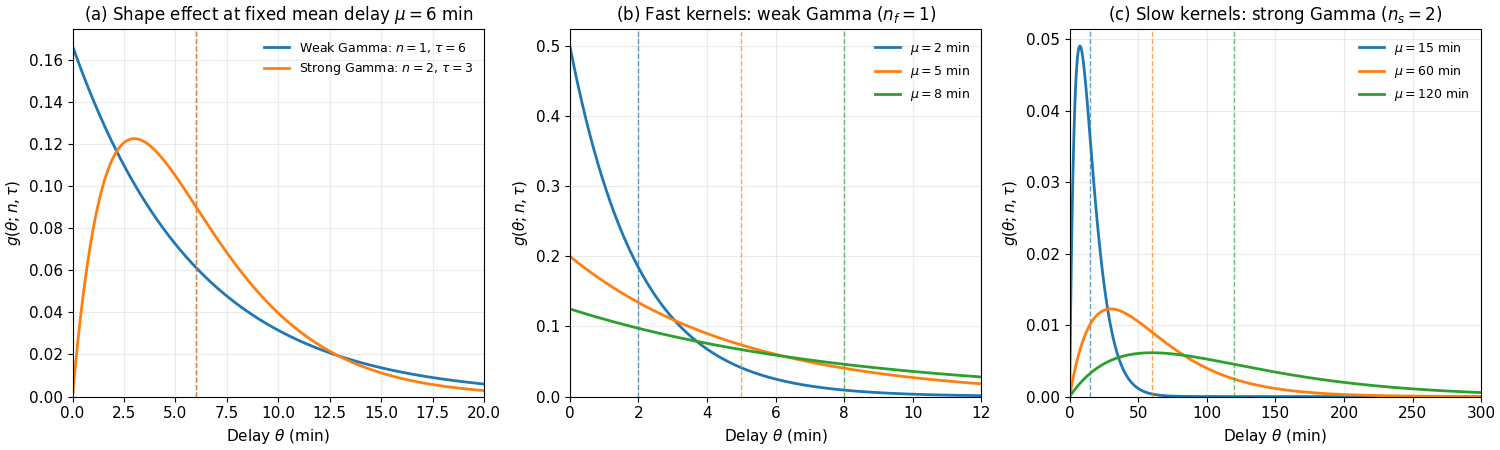}
\end{center}
\vspace{-5mm}
\caption{\small \emph{{\bf Weak and strong Gamma kernels for fast and slow stress modulation.}
(a) Comparison of weak ($n=1$) and strong ($n=2$) Gamma kernels at the same mean delay $\mu=6$ min. The weak kernel is maximal at zero delay, whereas the strong kernel has a delayed maximum and a smaller
variance at the same mean delay. For a Gamma kernel with order $n$ and mean $\mu$, the variance is $\mu^2/n$.
} (b) Weak Gamma kernels ($n=1$) for $\mu \in\{2,5,8\}$ min. (c) Strong Gamma kernels  ($n=2$) for $\mu \in\{15,60,120\}$ min. Dashed lines mark mean delays $\mu=n\tau$.}
\end{figure}

From a dynamical perspective, we ask how variation in the stress-sensitive couplings $w_{PP}$ and $w_{CB}$ shapes the behaviors produced by the model, in particular the transition between stable equilibria and sustained oscillations, and the associated changes in period, amplitude, duty cycle, and persistence. We then examine whether introducing cortisol-mediated delays, and changing how this delayed influence is distributed in time, shifts the parameter ranges over which these behaviors occur or changes the mechanisms by which oscillations appear and disappear. Biologically, these oscillations can be interpreted as slow alternations between periods of heightened limbic arousal and reactivity and periods of greater inhibition and regulatory control. Their period and duty cycle therefore reflect not only how often these transitions occur, but also how long the system remains in a more reactive versus a more suppressed state. The delay analysis therefore asks whether the timing and memory of cortisol action can reshape not only the level of activity, but also the temporal organization and persistence of cortico-limbic dynamics.

\subsection{Behavior of the system with no cortisol feedback}

We first analyze the neural system with constant coupling weights and without cortisol-related modulation.
We refer to this formulation as the cortisol feedback-free model, which provides a reference for the intrinsic dynamics of
the circuit.  In particular, this baseline analysis allows us to determine how the two stress-sensitive couplings ($w_{PP}$ and $w_{CB}$) balance one another in organizing stable and oscillatory regimes. We vary each parameter separately, then examine their joint effect in the $(w_{PP},w_{CB})$ plane.

\begin{landscape}
\begin{figure}[h!]
\begin{center}
\includegraphics[width=0.85\linewidth]{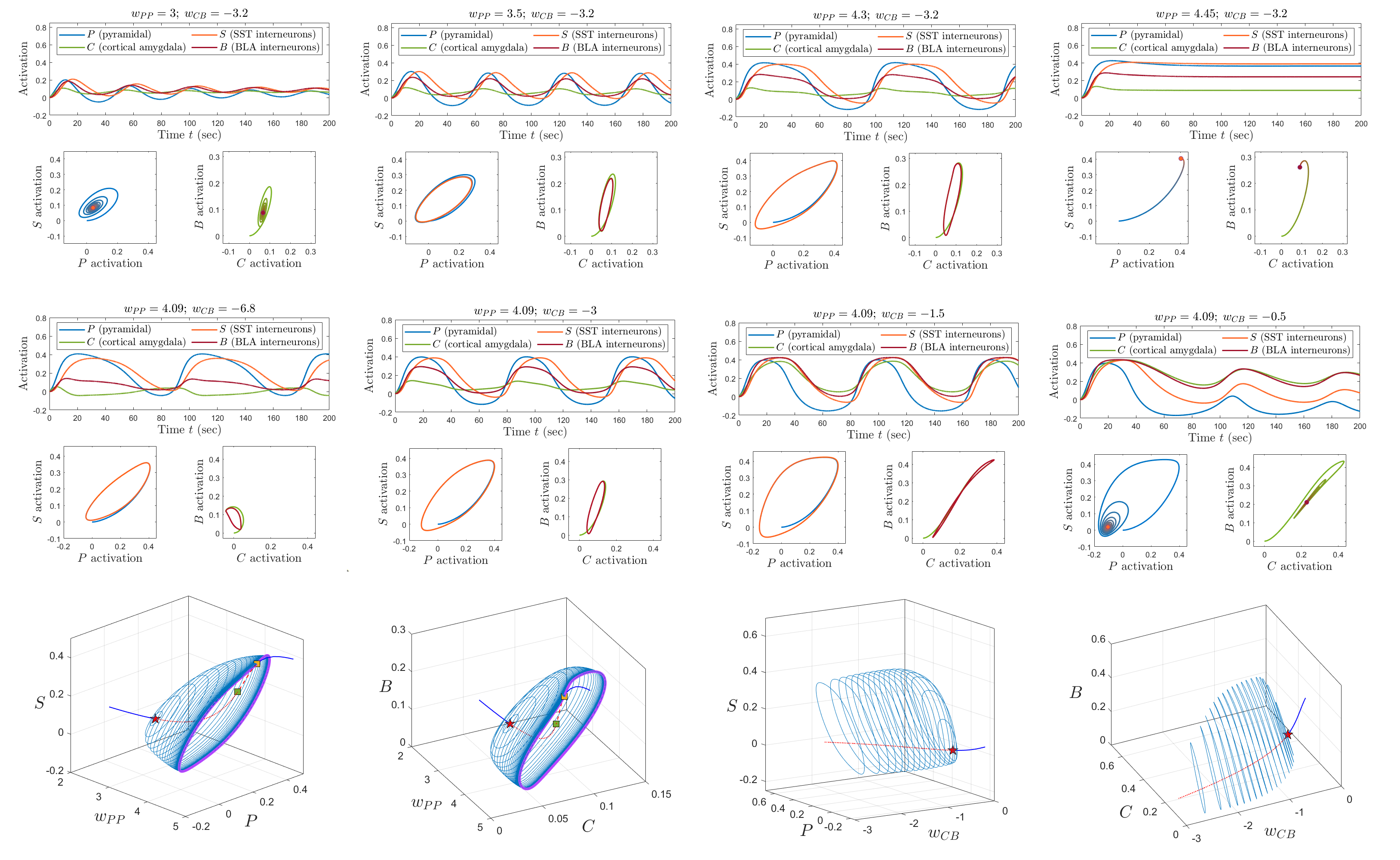}
\end{center}
\caption{\emph{{\bf Bifurcations with respect to $w_{PP}$} for fixed $w_{CB}=-3.2$, for the system with no time-delays. Subfigures A-D each show a temporal solution of the system for one value of the recurrent excitatory pyramidal coupling $w_{PP}$, as specified in the legends. The solution is illustrated in two ways: all system components with respect to time (top panels) and as a curve in the $(P,S)$ (prefrontal) and $(C,B)$ (amygdala) phase slices (bottom panels). The 3-dimensional plots E-F show the bifurcation diagram with respect to the key parameter $w_{PP}$, in a $(P,S)$ and respectively $(C,B)$ coordinate system. {\bf Bifurcations with respect to $w_{CB}$} for fixed $w_{PP}=4.09$, for the system with no time-delays. Subfigures A-D each show a temporal solution of the system for one value of the inhibitory amygdala coupling $w_{CB}$, as specified in the legends. The solution is illustrated in two ways: all system components with respect to time (top panels) and as a curve in the $(P,S)$ (prefrontal) and $(C,B)$ (amygdala) phase slices (bottom panels). The 3-dimensional plots E-F show the bifurcation diagram with respect to the key parameter $w_{CB}$, in a $(P,S)$ and respectively $(C,B)$ coordinate 
system.}}
\label{solutions_nodelay}
\end{figure}
\end{landscape}

In Figure~\ref{solutions_nodelay} we first illustrate the effect of increasing $w_{PP}$ for a fixed value of the 
inhibition parameter, $w_{CB}=-3.2$. This allows us to isolate how recurrent 
cortical excitation shapes the dynamics of the system when the inhibitory coupling is at a mid-range baseline level. For smaller values of 
$w_{PP}$, the trajectory approaches a stable equilibrium, indicating that 
recurrent excitation is not yet sufficient to sustain cycling. As $w_{PP}$ 
crosses the lower transition point, this equilibrium loses stability and the 
system enters an oscillatory regime. Near $w_{PP}=3$, the resulting oscillations 
are relatively small in amplitude and remain close to the former equilibrium. As $w_{PP}$ is increased further, the oscillations grow in amplitude and their 
temporal structure changes. At $w_{PP}=3.5$, the system exhibits a clear periodic 
orbit, with a more visible alternation between low- and high-activity phases. By 
$w_{PP}=4.3$, the oscillations become strongly relaxation-like: the active phase takes a larger fraction of the cycle, while the return to the low-activity 
phase is comparatively fast. Thus, increasing $w_{PP}$ alters not only the 
amplitude of the oscillation, but also its duty cycle and frequency. Near the upper end of the oscillatory window, the period increases substantially 
as the system approaches the transition out of the oscillatory regime. At 
$w_{PP}=4.45$, cycling has disappeared and the trajectory approaches a stable 
elevated equilibrium. This suggests that the loss of oscillations at high 
$w_{PP}$ occurs through a mechanism associated with progressive lengthening of 
the cycle, rather than through a simple small-amplitude decay. In this sense, 
increasing recurrent cortical excitation first destabilizes the equilibrium and 
generates oscillations, then produces larger and slower relaxation-like cycles, 
and finally drives the system into persistent high activation.

To connect the individual simulations above into a single contiguous picture, 
we performed a one-parameter bifurcation analysis in MatCont with $w_{CB}=-3.2$ 
fixed and $w_{PP}$ varied. The bottom panels plot the paired excitatory-inhibitory 
coordinates against $w_{PP}$, with the amygdala pair $(C,B)$ on the left and the 
cortical pair $(P,S)$ on the right. These projections show how the stable 
equilibrium branch gives way to sustained cycling as $w_{PP}$ enters the 
oscillatory window, how the periodic orbits expand with increasing recurrent 
cortical excitation, and how the oscillations terminate near the upper edge of 
the window. The upper termination of the periodic orbit has a SNIC-like geometry 
(saddle-node on invariant circle): the cycle appears to collide with the 
saddle-node point marked by the orange square, and this is accompanied by a 
progressive lengthening of the oscillation period. Although the MatCont 
continuation algorithm does not identify this transition as such, the observed 
period blow-up suggests a SNIC-mediated loss of cycling, after which the 
system settles onto a stable elevated equilibrium.

We next fix $w_{PP}=4$ and vary $w_{CB}$ to examine how changes in inhibitory 
control within the amygdala shape the dynamics when recurrent cortical excitation 
is held fixed to a mid-range level. Figure~\ref{solutions_decreasing_wCB} shows that, for weak inhibitory 
strengths, such as $w_{CB}=-0.2$ and $w_{CB}=-0.5$, the system does not sustain 
oscillations. Instead, the trajectories approach a stable elevated equilibrium, 
with the amygdala variables remaining at relatively high activation. As $w_{CB}$ 
becomes more negative, the system enters an oscillatory regime. At $w_{CB}=-1.5$, 
the model exhibits sustained cycling with large swings in both the cortical 
and amygdala variables. For still more negative values, such as $w_{CB}=-4.5$, 
the oscillations persist, although amygdala amplitudes are reduced compared with 
the less inhibited case. Hence, strengthening this inhibitory pathway can move the system away from persistent elevated activation through oscillatory regimes with different amplitudes and duty cycles. The bottom panels of Figure~\ref{solutions_decreasing_wCB} summarize this 
transition using a one-parameter bifurcation analysis in MatCont with $w_{PP}=4$ 
fixed and $w_{CB}$ varied. The projections connect the individual simulations 
into a contiguous bifurcation-level picture of the $w_{PP}=4$ parameter slice, 
showing how the stable equilibrium gives way to persistent stable cycling as 
$w_{CB}$ becomes more negative.

\begin{figure}[h!]
\begin{center}
\includegraphics[width=0.95\textwidth]{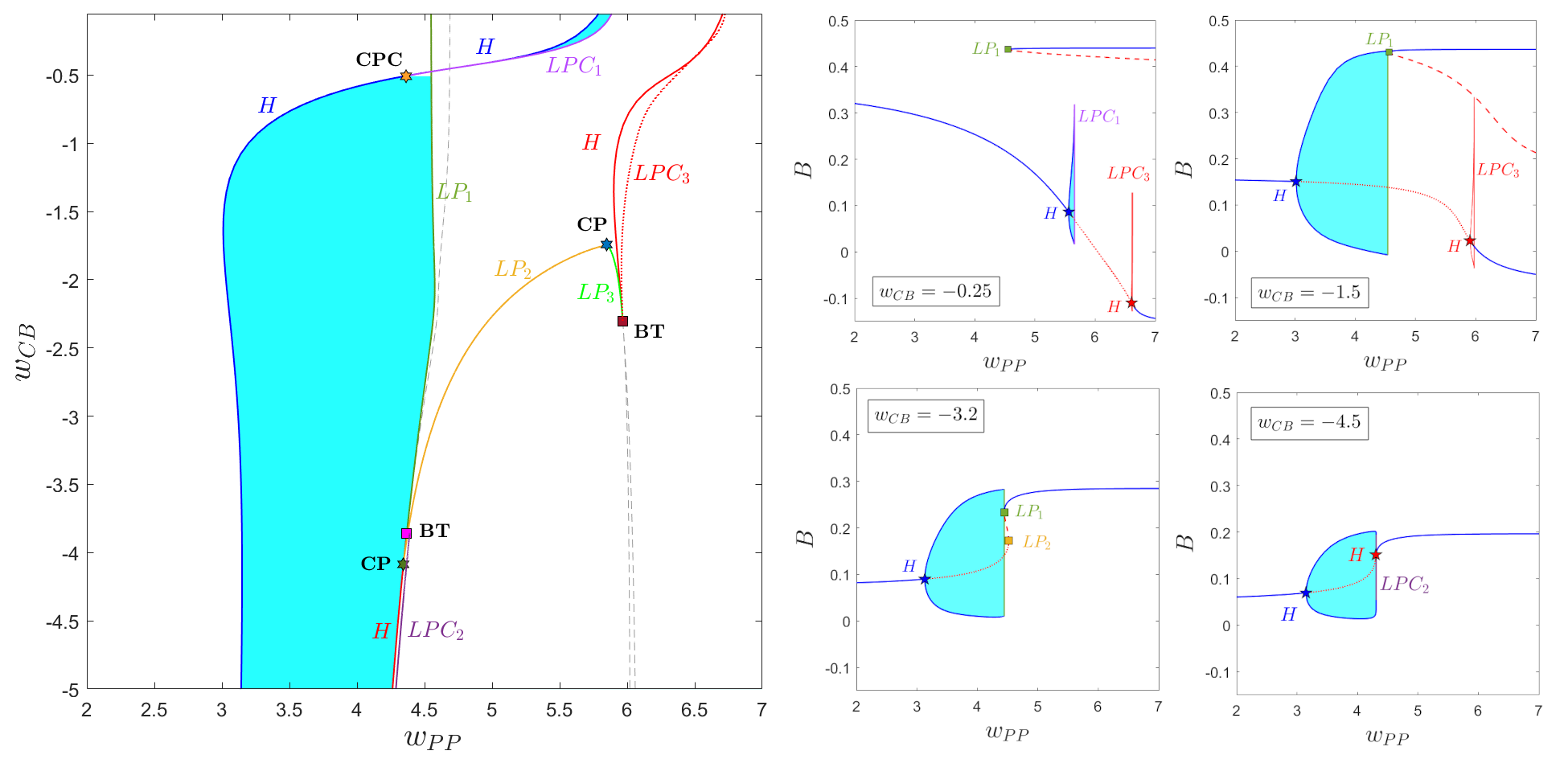}
\end{center}
\caption{ \emph{{Bifurcation curves in the key parameter-plane,  $(w_{PP},w_{CB})$,} for the system with no time-delays. A supercritical Hopf curve ($H$, shown in solid blue) spans the whole range for $w_{CB}$, providing the main mechanism for oscillation onset. The limit point of cycles curve $LPC_1$ provides a prompt end of oscillations for weak inhibitory $w_{CB}$, and terminates into a cusp point of cycles ($CPC$, shown as a yellow star). The red subcritical Hopf curve marks where the equilibirum regains stability; the end of unstable oscillations is marked by a limit point curve ($LPC_3$, shown as dotted red). Three limit point curves ($LP_1$, $LP_2$ and $LP_3$) are shown respectively in dark green, orange and light green, and meet at two cusp points (labeled $CP$). Between $w_{CB}=-4$ and $w_{CB}=-0.5$, $LPC_1$ contributes to the oscillation-ending mechanism, as shown in the side panels and discussed in the text. For inhibition lower than the cusp point at $w_{CB}=-4$, termination of cycles is provided by the limit point curve $LPC_2$ (shown in dark purple). The side panels illustrate one-parameter bifurcation slices with respect to $w_{PP}$, for four sample values of the inhibitory strength: $w_{CB}=-0.25$;  $w_{CB}=-1.5$; $w_{CB}=-3.2$ and $w_{CB}=-4.5$.}}
\label{diagram_nodelay}
\end{figure}

These preliminary explorations give an initial sense of the regimes and 
transitions that appear as the key parameters $w_{PP}$ and $w_{CB}$ are varied 
separately. To obtain a fuller picture of how these regimes are organized and 
jointly tuned by the recurrent cortical excitation and amygdala inhibitory 
strengths, we compute a two-parameter bifurcation diagram in 
the $(w_{PP},w_{CB})$ plane, shown in Figure~\ref{diagram_nodelay}. The bifurcation curves and codimension-two points 
are color-coded as described in the caption. The cyan shaded region indicates 
parameter values for which the system supports stable cycling activity. The 
diagram shows that oscillations are confined to two bounded regions of the 
$(w_{PP},w_{CB})$ plane. Overall, the diagram shows that low values of recurrent cortical excitation $w_{PP}$  are insufficient to destabilize the equilibrium, and the system remains non-oscillatory. As $w_{PP}$ increases, the system enters oscillations through the supercritical Hopf curve shown in blue. This oscillatory window is terminated within the observed parameter range, for all values of $w_{CB}$, with restabilization of the system to an equilibrium.

For weaker inhibition, corresponding to the top of the bifurcation diagram, the oscillatory window is narrowly ended by a limit point of cycles ($LPC_1$, purple curve). A sample section for $w_{CB}=-0.25$, presented in the first right panel, confirms that oscillations occur over a very brief $w_{PP}$ window; once stable cycles are lost through crossing the $LPC$ curve, the system will converge to the locally attracting equilibrium branch born from the limit point $LP_1$. This branch provides the only attractor until the original equilibrium regains stability through a subcritical Hopf bifurcation. This interval for $w_{PP}$ is flanked by two bistability windows over which the equilibrium branches coexist as stable attractors.

Stronger amygdala inhibition, corresponding to more negative values of $w_{CB}$, 
broadens the oscillatory region over a substantial interval of $w_{PP}$, with the 
inhibitory coupling $w_{CB}$ further shaping the cycling termination mechanism, as illustrated in both the $(w_{PP},w_{CB})$ plane as well as in the side panels for $w_{CB}=-1.5$ and $w_{CB}=-3.2$. For these mid-range values of $w_{CB}$, oscillations end via the SNIC-like transition: the cycle approaches a saddle-node structure, with blowup of the period before the cycle is lost. For stronger $w_{CB}$ (at the bottom of the diagram), oscillations end via a limit point of cycles. Interestingly however, the recurrent excitation level $w_{PP}$ at which oscillations terminate remains relatively unchanged across the board, irrespective of the termination mechanism. 

Taken together, these results establish the baseline dynamical organization of 
the system in the absence of temporal memory. Across the explored parameter 
region, oscillations emerge primarily through a supercritical Hopf bifurcation, 
while the mechanism by which they terminate depends more strongly on the level 
of amygdala inhibition. This baseline organization provides the reference point for examining how the introduction of cortisol-mediated temporal memory alters the system dynamics.

%%%%%%%%%%%%%%%

\subsection{Behavior with weak Gamma distributed delays}

\begin{landscape}
\begin{figure}[h!]
\begin{center}
\includegraphics[width=0.85\linewidth]{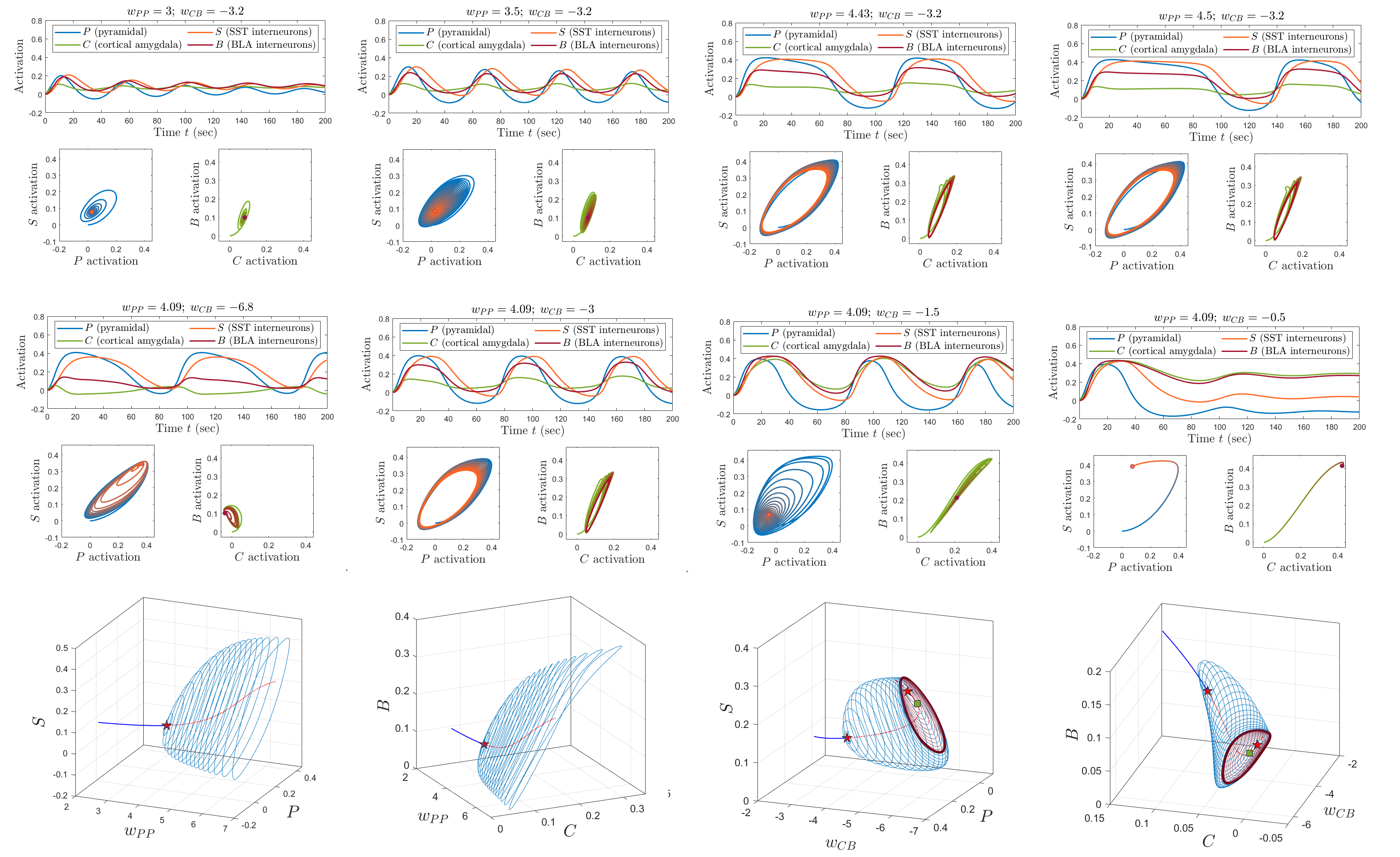}
\end{center}
\vspace{-7mm}
\caption{\emph{{\bf Bifurcations with respect to $w_{PP}$} for fixed $w_{CB}=-3.2$. Sub-figures A-D each show a temporal solution of the system for one value of the recurrent excitatory pyramidal coupling $w_{PP}$, as specified in the legends. The solution is illustrated in two ways: all system components with respect to time (top panels) and as a curve in the $(P,S)$ (prefrontal) and $(C,B)$ (amygdala) phase slices (bottom panels). The 3-dimensional plots E-F show the bifurcation diagram with respect to the key parameter $w_{PP}$, in a $(P,S)$ and respectively $(C,B)$ coordinate system.}}
\label{weak_solutions}
\end{figure}
\end{landscape}

We next redo the analysis for the model with weak Gamma distributed delays. For a proof of concept, we took $\tau_f=120$s and $\tau_s=1000$s, which are within the baseline physiological range for the fast and slow average delays, as specified in Table 1. To identify stable equilibria for our system with two (fast-slow) time-scales, we started with $w_{PP}=w_{CB}=0$ (where the delays have no impact), and extended from there, to avoid spurious results from direct integration in a stiff system. As a first experiment outlining differences in behavior between the weak delay system and the no delay system, we simulated numerically the solutions for the same parameter schemes as in Figure~\ref{weak_solutions} (with the system initiated at rest, i.e., at zero in all components). 

To do so, we first fixed the inhibition to $w_{CB}=-3.2$, and allowed $w_{PP}$ to change through the same values as in the right panels of Figure~\ref{solutions_nodelay}. Compared with the no-delay case, the transition into sustained cycling occurs at a higher value of recurrent prefrontal excitation. For $w_{PP}=3$ and $w_{PP}=3.5$, the trajectories show damped oscillatory transients that approach a stable equilibrium, rather than a stable periodic orbit. Sustained oscillations appear only after $w_{PP}$ is increased further. By $w_{PP}=4.43$, the system has entered a large-amplitude oscillatory regime, with a clear separation between low- and high-activity phases. The oscillations remain organized at $w_{PP}=4.5$, suggesting that weak Gamma delays shift the onset of cycling to larger $w_{PP}$ while also supporting robust large-amplitude oscillations once the transition has occurred. The bottom panels connect these simulations through a one-parameter MatCont continuation with $w_{CB}=-3.2$ fixed and $w_{PP}$ varied. The paired projections $(P,S)$ and $(C,B)$ show the stable equilibrium branch giving way to a stable cycle, which evolves as $w_{PP}$ increases. The oscillations extend over a broader range of $w_{PP}$ than in the case of no delays, indicating that weak temporal filtering can change not only the location of the transition but also the persistence of the oscillatory regime.

We then fixed $w_{PP}=4.09$ and varied $w_{CB}$, to compare how amygdala inhibition shapes the dynamics between the no-delay and weak-delay systems. For weak inhibition, $w_{CB}=-0.5$, both models approach a stable elevated equilibrium, although the delayed system displays a longer transient before settling. As $w_{CB}$ becomes more negative, both systems enter a large-amplitude oscillatory regime: at $w_{CB}=-1.5$ and $w_{CB}=-3$, the temporal profiles and phase-plane loops show sustained periodic cycling in both cases, while the weak-delay model exhibits more pronounced spiral transients toward the attracting cycle. At $w_CB=-6.8$, relaxation-like oscillations are still present in both models. The main distinction is revealed by the continuation diagrams: in the no-delay system, stable cycles persist as inhibition is strengthened across the explored range, whereas in the weak-delay system the periodic branch eventually terminates via a Homoclinic bifurcation at a finite value of strongly negative $w_{CB}$. 

\begin{figure}[h!]
\begin{center}
\includegraphics[width=\textwidth]{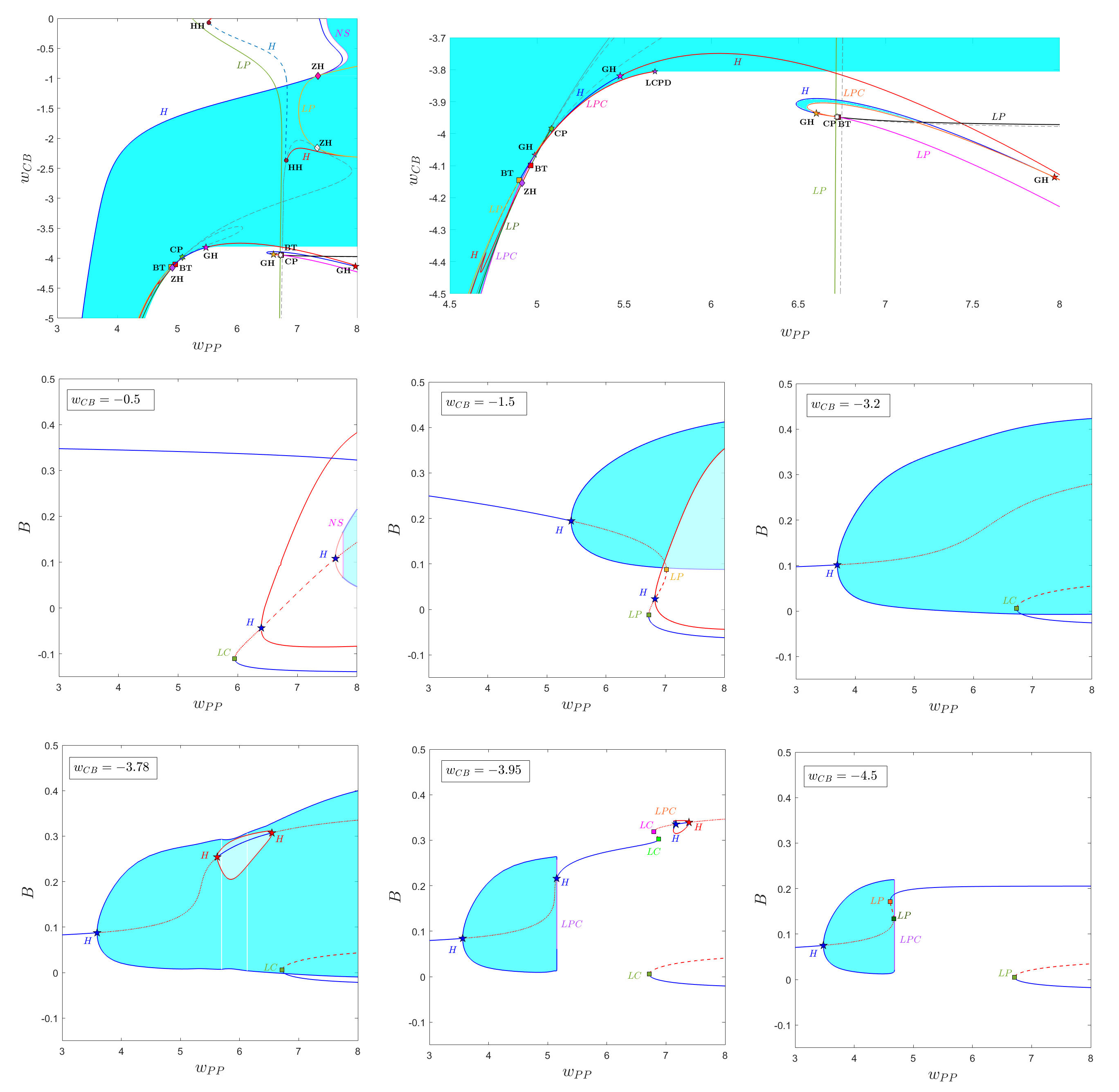}
\end{center}
\caption{\emph{{\bf Two-parameter bifurcation structure for the weak Gamma delay model in the $(w_{PP},w_{CB})$ plane.} The cyan region marks parameter values for which the system supports stable periodic activity. Hopf ($H$), limit-point ($LP$), and limit-point cycle ($LPC$) bifurcation curves organize the onset and loss of sustained oscillations; neutral saddle curves are included for completeness (dotted lines, $NS$) when they affect nearby stability structure, but do not themselves delimit oscillatory regimes. {\bf A.} Main bifurcation curves in the full parameter window that was investigated. {\bf B.} Magnification of the complex bifurcation structure and oscillation boundary in the region $w_{CB} \in [-4.5,-3.5]$. Hopf curves are labeled $H$. There are three supercritical Hopf curves (shown in blue), and three subcritical Hopf curves (shown in red). There are six limit point curves: two prominent ones (green and orange) are labeled in panel (A), and the additional ones are only labeled in the magnified panel (B), to avoid cluttering. Codimension-two points are labeled as follows: cusp points (CP), generalized Hopf points (GH), Bogdanov--Takens points (BT), zero-Hopf points (ZH), Hopf-Hopf points (HH) and limt cycle period doubling ($LCPD$). {\bf (C)-(H).} To better illustrate the behaviors encountered, these panels show representative one-parameter sections with $w_{CB}$ fixed at six increasingly stronger levels of inhibition, plotting in each case the equilibrium branch and cycles as $w_{PP}$ varies.}}
\label{diagram_weak}
\end{figure}

The two-parameter diagram in Figure~\ref{diagram_weak} classifies these behaviors for the functional parameter region $(w_{PP},w_{CB}) \in [3,8] \times [-5,0]$, showing that weak Gamma filtering substantially reorganizes the bifurcation structure in the $(w_{PP},w_{CB})$ plane. As in the no-delay model, the main attracting behavior is organized around a region of stable periodic solutions, shown in cyan. As $w_{PP}$ is increased from zero, the system always enters this region, as before, through a supercritical Hopf bifurcation; however, the shape of this curve, as well as fundamentally different options for ending the cycles, give this oscillatory region a different shape and meaning than for the system in absence of delays.

The figure illustrates this supercritical Hopf boundary (blue curve), as well as all other meaningful bifurcation curves that contribute to the shaping of the transitions in and out of oscillations, including other supercritical and subcritical Hopf curves, limit point curves, limit points of cycles, as well as neutral saddle curves for completeness. Some of the most relevant codimension-2 bifurcation points are also shown, to emphasize the intersections of these curves and assemble an overall map of the system's phase transitions. Figure 8a portrays these bifurcations in the whole parameter region of interest $[3,8] \times [-5,0]$. Figure 8b shows a blowup of the mid $w_{CB}$ region, where a bundle of bifurcations occur in a very tight space (as we explain below), including two supercritical Hopf curves with onset of stable cycles. These cycles are ended in both cases by limit points of cycles (LPC, or fold bifurcations). To avoid over-cluttering, the LPC curves are omitted in panel (a), but are shown in the zoomed-in version in panel (b), as part of the boundary for stable oscillations. To clarify this bigger picture, panels (c)-(h) show one-parameter sections with $w_{CB}$ fixed to six different representative values,  illustrating the equilibria and cycles, and breaking down how the system enters and leaves the oscillatory window in each case.

Notice that, for weaker inhibition (i.e., values of $w_{CB}$ closer to zero, corresponding to the top part of the parameter plane) the onset of oscillations is shifted to larger values of the recurrent prefrontal excitation $w_{PP}$. However, for this parameter range, the weak Gamma memory does not merely shift the Hopf onset in this inhibition range; it also changes the persistence of the oscillatory state. The stable cycle formed here persists across the remainder of the explored $w_{PP}$ interval (visible in all one-parameter sections $w_{CB}=-1, -1.5, -3.2$ and $-3.78$). Simultaneously, the unstable equilibrium branch after the Hopf bifurcation will undergo further changes in stability as it crosses additional limit points and subcritical Hopf points. This creates a bistability window with coexistence of a stable equilibrium and a stable cycle, which persists for roughly half of the $w_{CB}$ domain (present in different forms in all sections from $w_{CB}=-1$ to $w_{CB}=-3.78$). 

The situation changes for strong inhibition $w_{CB}$, where the oscillatory region starts at lower $w_{PP}$ values than in the no delay model, and it is also cut short by a limit point of cycles ($LPC$ curve) at lower values of $w_{PP}$. Indeed, the sections for $w_{CB}\approx -3.95$ and $w_{CB}=-4.5$ confirm that stable cycles exist only over a finite $w_{PP}$ interval before disappearing. For inhibitory values around $w_{CB}=-4$, the stable oscillations reappear for an extremely short excitation interval ranging between $w_{PP}=6-7$, delimited by a supercritical Hopf curve, and promptly stopped by a corresponding $LPC$ curve (shown on the right side of the zoom-in panel, and further illustrated in the one-parameter section corresponding to $w_{CB}=-3.95$). This secondary oscillatory region was included in the diagram for completion, but is small and confined to a very limited range of $w_{CB}$, so we will exclude it from our further discussions of oscillations and transitions. Also notice that, for inhibition stronger than $w_{CB}=-4$, the equilibrium-cycles bistability windows are negligible, and are replaced by the possibility of two coexisting equilibria, if both inhibition $w_{CB}$ and excitation $w_{PP}$ are strong (see for example the right side of the plot in panel (H)).

One can note that, broadly speaking, the boundary of the oscillatory region for weak Gamma delays is organized by a much richer collection of local and global bifurcations than in the no-delay model. Hopf curves still provide the main mechanism for the entry into oscillations, but generalized Hopf, Bogdanov--Takens, zero-Hopf, Hopf-Hopf, cusp, and limit-cycle bifurcation clusters are now abundant, especially in the $w_{CB}$ mid-range of the diagram. This richness indicates that the weak-delay terms introduce an additional layer of complexity in the mechanisms that control cycling behavior, emergence of stable equilibrium branches and bistability. For the purposes of the present interpretation, the essential point is not the identity of every local transition, but the global effect: weak Gamma memory reshapes and expands the parameter region in which stable cycling can persist with increasing $w_{PP}$, while also creating a tighter excitatory cutoff at sufficiently negative $w_{CB}$.

\clearpage
On its own, the weak-delay diagram shows that history-dependent cortisol modulation delays the onset of oscillations in parameter space, shifting it to larger $w_{PP}$, but removes the upper termination of the oscillatory regime within the explored $w_{PP}$ range. The delay variables act like low-pass filtered versions of recent amygdala activity, so the circuit no longer responds only to the instantaneous balance of prefrontal recurrence and amygdala inhibition. Instead, it responds to a short memory of recent activation. That memory can initially stabilize the equilibrium, which explains why larger $w_{PP}$ is needed to initiate stable cycling. But once the cycling state is reached, the delayed feedback helps maintain the oscillation between low- and high-activity phases over a much wider parameter range.

Biologically, this is a useful interpretation: weak cortisol memory does not simply “increase oscillations.” It changes the conditions under which oscillations are born and the conditions under which they die. Moderate inhibition allows the system to enter a persistent rhythmic regime, which can be interpreted as repeated alternation between more regulated and more reactive cortico-limbic states. But excessive inhibition can suppress this cycling, producing a threshold beyond which the system no longer supports sustained state transitions.

\subsection{Behavior with strong Gamma distributed delays}

For the strong-Gamma model, we set $n_f=n_s=2$, $\tau_f=60$ s and $\tau_s=500$ s. These scale parameters
preserve the mean delays $\mu_f=n_f\tau_f=120$ s and $\mu_s=n_s\tau_s=1000$ s
used in the weak-Gamma model.
The comparison therefore changes the kernel shape while keeping both mean delays fixed. This  will allow us to directly compare the effect of how the kernel shaping the dependence of past arousal cortisol impacts future dynamics, even for the same average delays.

\begin{figure}[h!]
\begin{center}
\includegraphics[width=\textwidth]{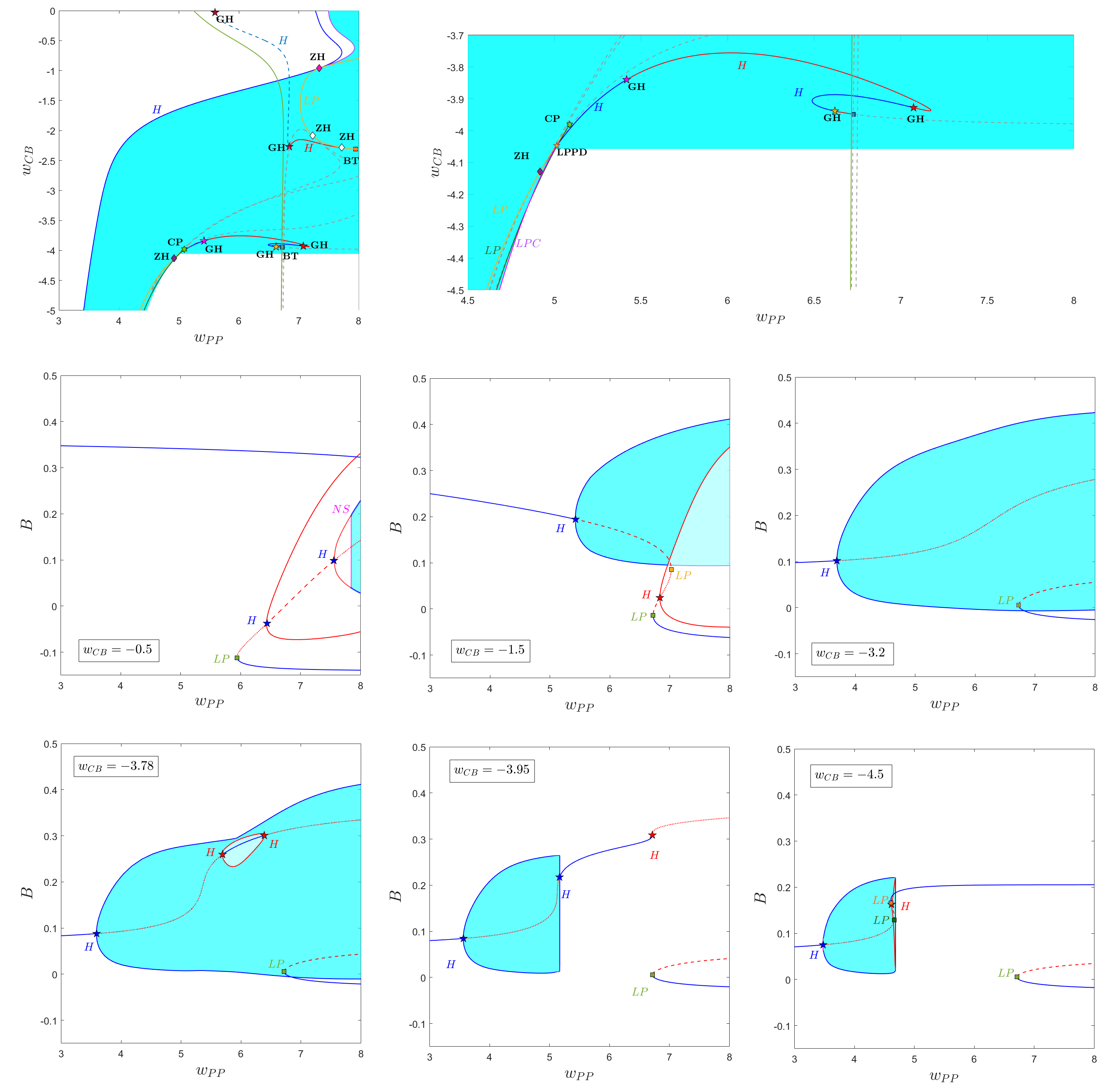}
\end{center}
\caption{\emph{{\bf Two-parameter bifurcation structure for the strong Gamma delay model in the $(w_{PP},w_{CB})$ plane.} The cyan region marks parameter values for which the system supports stable oscillations. Hopf ($H$), limit-point ($LP$), and limit-point cycle ($LPC$) bifurcation curves organize the onset and loss of sustained oscillations; neutral saddle curves are included for completeness (dotted lines, $NS$) when they affect nearby stability structure, but do not themselves delimit oscillatory regimes. {\bf A.} Main bifurcation curves in the full parameter window that was investigated. {\bf B.} Magnification of the complex bifurcation structure and oscillation boundary in the region $w_{CB} \in [-4.5,-3.5]$. Hopf curves are labeled $H$. There are three supercritical Hopf curves (shown in blue), and three subcritical Hopf curves (shown in red). There are six limit point curves: two prominent ones (green and orange) are labeled in panel (A), and the additional ones are only labeled in the magnified panel (B), to avoid cluttering. Codimension-two points are labeled as follows: cusp points (CP), generalized Hopf points (GH), Bogdanov--Takens points (BT), zero-Hopf points (ZH), Hopf-Hopf points (HH) and limt cycle period doubling ($LCPD$). {\bf (C)-(H).} To better illustrate the behaviors encountered, these panels show representative one-parameter sections with $w_{CB}$ fixed at six increasingly stronger levels of inhibition, plotting in each case the equilibrium branch and cycles as $w_{PP}$ varies.}}
\label{diagram_strong}
\end{figure}

We found that the change from weak to strong Gamma kernel in the delay distribution triggered only minor adjustments in the bifurcation map, leaving the major transitions in and out of cycling behavior virtually unaltered. The corresponding diagrams are shown in Figure~\ref{diagram_strong}, for comparison. The near coincidence of the weak- and strong-Gamma bifurcation diagrams provides evidence for a local robustness of the dynamics with respect to the shape of the distributed memory kernel. At fixed mean delay, increasing the Gamma order from $n=1$ to $n=2$ changes how past activity is weighted: the kernel becomes less concentrated near the present and more strongly organized around a delayed peak. Despite this change, the principal bifurcation curves, the oscillatory region, and the mechanisms governing the onset and termination of cycling remain almost unchanged. Thus, the bifurcation structure is not tied to one particular low-order Gamma kernel. At the same time, this comparison cannot establish robustness across the full family of Gamma distributions: the passage from $n=1$ to $n=2$ represents only one step toward increasingly concentrated memory, and a systematic continuation over all kernel orders is beyond the scope of the present study.

Biologically, this suggests that the cortico-limbic consequences of cortisol-mediated memory are buffered against moderate variation in the temporal spread of cortisol action. Such variation is expected, since receptor kinetics, intracellular signaling, and downstream synaptic effects can differ across individuals and physiological conditions. The similarity between the weak- and strong-Gamma cases therefore indicates that modest changes in the breadth and timing of this distributed processing do not fundamentally alter the predicted circuit response. However, this does not imply that the temporal organization of cortisol action is unimportant. Both kernels still represent broadly distributed processing of past amygdala activity, and they differ only in degree. To probe whether this apparent robustness persists under a much sharper form of temporal selection, we next consider the limiting case $n \to \infty$, in which a Gamma kernel with fixed mean collapses toward a discrete delay. Rather than examining every intermediate value of $n$, this comparison allows us to test whether the qualitative dynamics remain unchanged at the extreme limit of increasingly concentrated memory.

\clearpage
\subsection{Behavior with discrete delays}

We finally consider the limiting case in which the distributed memory is replaced by discrete delays,
\begin{eqnarray*}
w_{PP}(t) &=& w_{PP}[1-\beta_{PP} C(t-\tau_s)]\\
w_{CB}(t) &=& w_{CB}[1-\beta_{CB} C(t-\tau_f)]
\end{eqnarray*}

\noindent using the same working delay values as in the distributed-delay analysis, $\tau_f=120$ and $\tau_s=1000$. The corresponding delay differential equations were analyzed in Matlab, using the Delay Equation Importer implemented in MatCont 7p6. The importer approximates the infinite-dimensional delay system by a finite-dimensional system using a pseudospectral representation of the solution history over the delay interval, allowing the standard continuation and bifurcation algorithms implemented in MatCont to be applied. For the computations presented here, we used a pseudospectral approximation of order $M=10$. Thus, unlike the Gamma-distributed formulations, where the distributed delays can be rewritten exactly as finite-dimensional systems through the linear-chain construction, the discrete-delay formulation approximates the corresponding history numerically.

\begin{figure}[h!]
\begin{center}
\includegraphics[width=\textwidth]{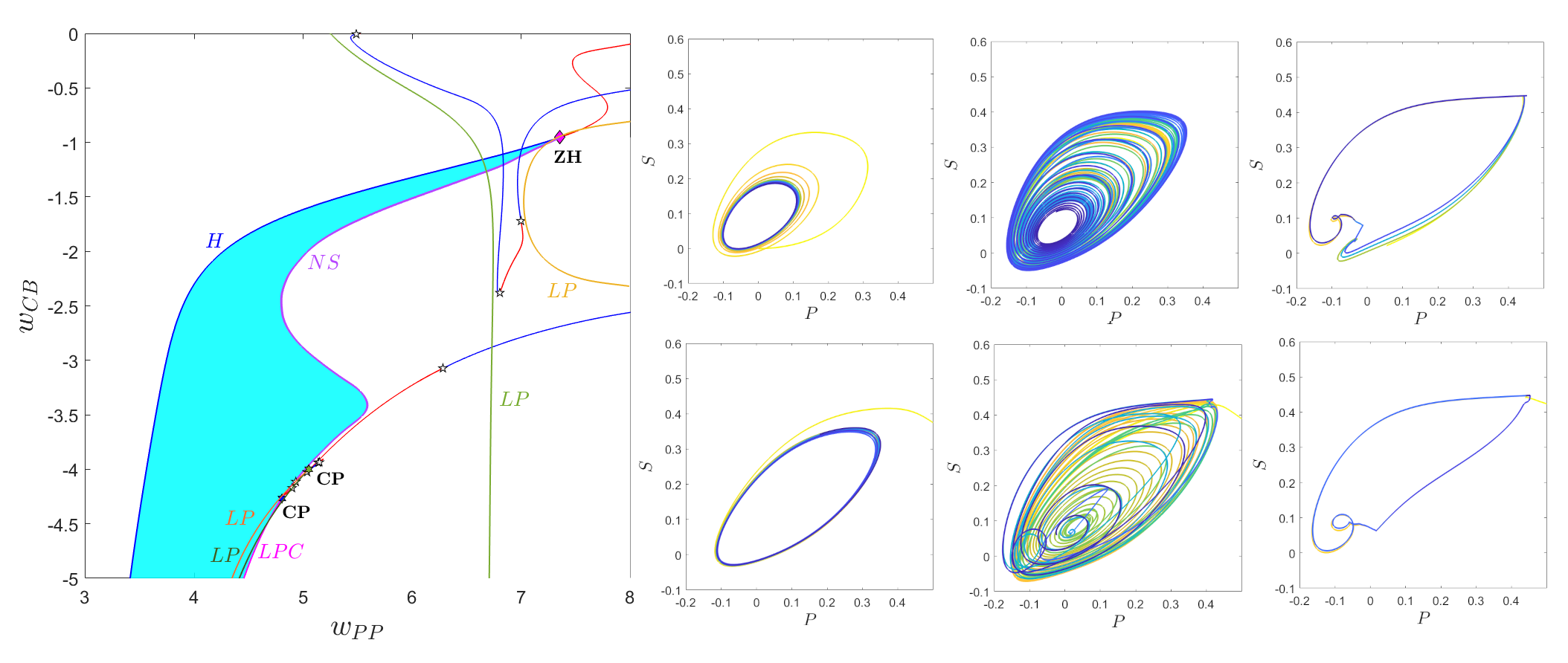}
\end{center}
\caption{\emph{{\bf Two-parameter bifurcation structure for the discrete delay model in the $(w_{PP},w_{CB})$ plane.} The cyan region marks parameter values for which the system supports stable oscillations. Hopf ($H$), limit-point ($LP$), limit-point cycle ($LPC$) and Neimark-Sacker ($NS$) bifurcation curves organize the onset and loss of sustained oscillations. The main Hopf curve where stable oscillations are born is shown in blue and labeled $H$ (for its supercritical portion); it is shown in red after it becomes subcritical at a Zero-Hopf point (labeled $ZH$, shown as a pink diamond). Stable oscillations end in the upper region via a $NS$ curve (purple)  and in the lower region via a $LPC$ curve (pink). Four Limit Point curves are labeled $LP$ and shown in different colors, with cusp point marked as $CP$. Two additional Hopf branches are shown (with supercritical portions in blue, and subcritical branches in red, with the Generalized Hopf transitions marked as small white stars (and not labeled, to avoid clutter). Other Hopf curves that do not directly contribute to the relevant dynamics are not shown, to preserve clarity of the diagram.}}
\label{discrete_diagram}
\end{figure}

As shown in Figure~\ref{discrete_diagram}, the bifurcation structure becomes considerably richer when the distributed memory is replaced by discrete delays, because of the infinite characteristic spectrum. In contrast with the finite-dimensional systems obtained from the weak and strong Gamma kernels, the system with discrete delays produced a flurry of Hopf branches in the $(w_{PP},w_{CB})$ parameter plane. While in our numerical exploration we recorded each of these branches individually, here we are reporting and will focus only on the bifurcations that organize the attracting dynamics of interest, and use the two-parameter diagram to understand the surrounding structure. In particular, most Hopf curves and codimension-two bifurcation points which were detected during continuation, but do not contribute to this structure are not shown in the figure, to avoid excessive clutter.

Despite this increased complexity, the principal onset mechanism for stable cycling remains easily identifiable. As in the distributed-delay systems, increasing recurrent prefrontal excitation $w_{PP}$ carries the system across the supercritical Hopf branch shown in solid blue, where a stable cycle is created. This mechanism persists for all levels of amygdala inhibition up to $w_{CB} \sim -1$, after which stable oscillations are no longer sustainable. The region in which these periodic solutions remain stable is shown in cyan, and is bounded at the higher $w_{PP}$ end by a Neimark-Sacker curve (purple curve labeled $NS$), for weaker inhibition from $w_{CB}=-1$ to $w_{CB}=-4$ and by a limit point of cycles (pink curve labeled $LPC$), for inhibition that is stronger than $w_{CB}=-4$. Hence for discrete delays as well, depending on the strength of amygdala inhibitory coupling, the same stable periodic regime can lose stability through two qualitatively different mechanisms when recurrent prefrontal excitatory coupling is increased.

Other elements of the bifurcation architecture are reminiscent of the distributed-delay cases. In particular, a green transition curve analogous to that observed for the weak and strong Gamma delays is retained, together with several limit-point branches of equilibria. The main bifurcation panel also shows the Zero-Hopf ($ZH$) bifurcation where a $LP$ branch and the $NS$ curve intersect with the primary Hopf curve (at the switch between its supercritical and subcritical branches). We also illustrate two cusp ($CP$) points, and a few of the Bautin bifurcations ($GH$) where the Hopf curves switch type, without new stable cycling being formed or destroyed.

The extensive Neimark-Sacker curve is a significant change from the weak and strong Gamma figures, since it does not mark the end of the stable oscillatory behavior altogether, but rather separates two distinct attracting oscillatory regimes: stable periodic cycling on the cyan side of the boundary and stable quasiperiodic dynamics over at least part of the parameter region beyond it. MatCont does not have a direct procedure for subsequently following the invariant torus through numerical continuation, hence we did not attempt to determine directly the full extent of the torus stability region, or the mechanism by which it ultimately terminates. Instead, we illustrate in the side panels that the stable torus is still present (with different geometry and kinetics) at two snapshots of $w_{PP}$ past the NS curve, for two separate sample inhibition values $w_{CB}=-2$ and $w_{CB}=-3.2$, respectively.

What is particularly interesting in the discrete-delay case is the extent of this Neimark-Sacker boundary. Quasiperiodicity is no longer confined to a very narrow parameter range, but becomes relevant over a large portion of the explored $(w_{PP},w_{CB})$ domain. Thus, for a broad range of amygdala inhibition, loss of regular periodic cycling does not mean loss of cortico-limbic oscillations. Instead, the circuit remains oscillatory while the timing and shape of successive loops become modulated. The discrete delay therefore allows recurrent prefrontal excitation to destabilize the regular rhythm without simply suppressing oscillations altogether.

The side panels further show that the resulting quasiperiodic dynamics are themselves strongly affected by $w_{PP}$. As recurrent prefrontal excitation is increased, the trajectory takes substantially longer to complete a full loop around the torus. This slowing suggests progressively longer modulation time-scales in the interaction between prefrontal and amygdala populations: the circuit may continue to alternate between regulatory and reactive configurations, but the return to a comparable network state occurs over increasingly long intervals. Hence, in this regime, changes in recurrent prefrontal excitation affect not only whether the circuit oscillates, but also how regularly and how quickly cortico-limbic states are revisited.

\section{Discussion}

\subsection{Specific comments on the model}

In this work, we investigated the effect of distributed delays on prefrontal-limbic interactions, interpreting these delays as a phenomenological representation of the temporal modulation produced by cortisol-mediated processes during emotion regulation. Rather than treating cortisol as an independent dynamical variable or as a direct generator of neural oscillations, we modeled its influence through delayed, history-dependent changes in two functionally important circuit couplings (which we treated as key parameters of the system): the inhibitory control of basolateral amygdala principal cells, represented by the coupling weight $w_{CB}$, and recurrent excitation within the prefrontal pyramidal population, represented by $w_{PP}$. The first was used to represent relatively fast cortisol-related modulation of amygdala inhibition, while the second captured slower effects on recurrent prefrontal recruitment. In this formulation, the delay structure determines how past amygdala activity contributes to the current effective strengths of these pathways, and therefore how the recent history of circuit activation reshapes the conditions for transitions between stable equilibria and stable  oscillations.

We considered three different temporal representations of this history dependence. The first two were the simplest members of the Gamma family of distributed delays: a weak Gamma kernel ($n_f=n_s=1$), which places the greatest weight on the recent past and decays monotonically with time, and a strong Gamma kernel ($n_f=n_s=2$), for which the contribution of past activity is shifted away from the present and organized around a delayed maximum. Finally, we considered a discrete delay as the limiting case of increasingly concentrated Gamma distributions at fixed mean delay. This progression allowed us to distinguish effects that depend primarily on the presence of temporal memory from those that depend more specifically on how broadly that memory is distributed in time.

Our model predicts substantial changes to this picture once delayed cortisol action is introduced. The main changes were not specific to one particular form of temporal weighting. Across the distributed-delay cases, introducing a history of cortisol-related modulation (with average delay consistent with free cortisol courses) reorganized the parameter region supporting infra-slow oscillations and altered their persistence (with only the finer details of the dynamics depending on the shape of the delay kernel).
In particular, the onset of stable cycling was shifted toward stronger recurrent prefrontal excitation: larger $w_{PP}$ was required before the stable equilibrium lost stability and sustained oscillations emerged. Once established, however, the periodic regime became considerably more persistent with increasing $w_{PP}$. In the no-delay system, sufficiently strong recurrent excitation ultimately terminated cycling and drove the system toward a stable, amygdala-elevated equilibrium, whereas under weak Gamma memory stable oscillations persisted over a much broader range of recurrent excitation. The effect of amygdala inhibition was also altered. Moderate inhibitory strength supported the periodic regime, but sufficiently strong inhibition eventually suppressed it, introducing a lower cutoff in $w_{CB}$ that was not a dominant feature of the instantaneous system. 

This distinction suggests a useful interpretation of the delay-system dynamics. The parameters $w_{PP}$ and $w_{CB}$ in the bifurcation diagrams represent the baseline coupling strengths, while cortisol acts on their \emph{effective} values through temporally filtered representations of recent amygdala activity. Thus, for fixed baseline $w_{PP}$ and $w_{CB}$, the circuit does not experience fixed coupling strengths: the effective recurrent excitation and amygdala inhibition vary in time according to the recent history of activation. This history dependence can initially stabilize the equilibrium, so that stronger baseline recurrent prefrontal excitation is required to trigger sustained cycling. Once oscillations are established, however, the delayed modulation can transiently shift the effective couplings into ranges that continue to support cycling, allowing infra-slow oscillations to persist at baseline parameter values for which the instantaneous system would already have left the oscillatory regime. Cortisol-like memory therefore does not produce a simple monotone translation of the oscillatory region, but reshapes the region of baseline $(w_{PP},w_{CB})$ parameter space in which sustained cycling is possible.

For weak and strong Gamma delays, the system has automatically the same equilibria -- but there is generally no expectation for robustness of bifurcations and oscillatory regions and mechanisms. In our model however, quite surprisingly, changing the delay distribution from the weak to the strong Gamma kernel had very little effect on this global picture. At the same mean delays, the major bifurcation boundaries and the region supporting stable periodic activity remained almost unchanged. This is notable because the two kernels weight the history of amygdala activity differently: the weak Gamma kernel emphasizes the immediate past, whereas the strong Gamma kernel suppresses the contribution of the present and places greater weight around a delayed peak. Nevertheless, these differences in temporal weighting produced only minor changes in the principal transitions between attracting states. Within the range examined here, the model therefore appears locally robust to moderate changes in the shape of distributed cortisol memory. The principal dynamical effect arises from introducing history dependence itself, while the precise distribution of that history between these two low-order Gamma kernels plays a comparatively smaller role.

The discrete-delay case represents a qualitatively different limit, in which this temporal averaging is removed and the modulation depends on activity at a sharply specified time in the past. In this limiting case, the bifurcation structure became substantially richer. Although the principal stable periodic regime was still created through a supercritical Hopf bifurcation, its subsequent evolution differed qualitatively from that observed with distributed memory. Over the portion of the oscillatory boundary corresponding to weaker amygdala inhibition, the stable cycle lost stability through a Neimark-Sacker bifurcation, beyond which direct numerical simulations revealed stable invariant tori (while for stronger amygdala inhibition, stable cycles instead terminated through a limit point of cycles). The discrete-delay model therefore supports not only equilibrium and periodic attracting states, but also stable quasiperiodic dynamics on extensive parameter ranges. 

The contrast between the distributed and discrete cases is an interesting result of the model. Moving from the weak to the strong Gamma kernel changes the relative weighting of recent versus more remote activity but leaves the main dynamical organization remarkably stable. Concentrating that memory onto a single past time, by contrast, introduces qualitatively new dynamics. Distributed memory appears to smooth the delayed feedback sufficiently to preserve a robust periodic organization, whereas sharply localized feedback can introduce an additional temporal modulation of the existing cycle and produce quasiperiodic activity. Hence our results suggest that the dynamic consequences of delayed cortisol action may depend not only on its characteristic time scale, but also on whether its effects integrate neural activity over an extended temporal window or depend much more selectively on activity at a particular past time.

These results reflect the idea that delayed cortisol effects are not a static increase or decrease in coupling strength, but rather introduce \emph{history} into the effective connectivity of the circuit. Once the present values of the stress-sensitive couplings depend on previous activity, the conditions for entering, sustaining, and leaving an oscillatory state can change substantially. Incorporating temporal history reshapes the stability landscape and allows dynamical regimes that are not available to the corresponding instantaneous system. The form of that history nevertheless matters in refining the resulting dynamics, with discrete delays providing a richer dynamic atlas than weak or strong Gamma kernels, consistently with results from our previous modeling work~\cite{kaslik2018stability,kaslik2022stability,kaslik2024stability}. This suggests the possibility that there is no single temporal-integration profile  that is implemented across all neural systems: different kernels may effectively characterize different circuits, physiological states, or cognitive tasks. An integration profile that is well suited to one process may be maladaptive in another, shifting a circuit toward atypical or pathological rhythmic regimes. While this interpretation remains a hypothesis, it suggests that the shape of neural memory may be functionally important in its own right.

\subsection{General comments}

Physiologically, our results support a view of cortisol as a temporally structured modulator of cortico-limbic regulation, rather than as a signal that simply increases or decreases neural activity. Experimental studies establish that the neural and behavioral effects of cortisol change strongly with time, but these studies are intrinsically limited by their ability to sample this evolution only at a small number of discrete observation windows~\cite{pan2023difference,henckens2010time,henckens2011time,jentsch2019restoring}. In parallel, mathematical models of the HPA axis have long represented endocrine transport and feedback using distributed delays, including weak and strong Gamma kernels and their discrete-delay limit~\cite{kim2016onset,malek2015dynamics,churilov2022modeling}. Our work brings these two ideas together at the level of cortico-limbic regulation, using distributed temporal weighting to represent how the history of cortisol-related arousal modulates effective neural interactions~\cite{kaslik2018stability}.

We should first note that infra-slow oscillations in our model do not require delayed cortisol action. They are already present in the instantaneous system, in which cortisol-sensitive modulation depends on the current circuit state but has no access to its previous history. This is consistent with experimental evidence that slow rhythmic activity can arise in neural tissue even when it is decoupled from the ongoing endocrine system. Dissociated cortical networks generate network-wide infra-slow oscillations below 0.01 Hz \cite{mok2012ultra}, while parahippocampal slices containing basolateral/lateral amygdala exhibit spontaneous periodic activity recurring every 5-30 s under control conditions \cite{kano2005periodic}. These experiments support the idea that cortisol integration is not required for neural circuitry to support slow rhythmic dynamics, but instead modifies the conditions under which the cortico-limbic circuit expresses and sustains these oscillations.

Even in preparations decoupled from the endocrine system, the expression of infra-slow rhythmic activity depends strongly on the balance of local circuit interactions. In developing cortical slices, spontaneous network oscillations in the infra-slow range ($\sim 0.01$ Hz) depend on glutamatergic excitation, and are strongly reduced or abolished when excitatory glutamatergic transmission is blocked \cite{garaschuk2000large,allene2008sequential}. This is consistent with the lower boundary of the oscillatory region in our model, where recurrent prefrontal excitation must exceed a minimum strength before sustained cycling can emerge through a Hopf bifurcation. Once oscillations are established in our system, increasing $w_{PP}$ progressively lengthens the high-activity portion of the cycle, until the period blows up and the system ceases to return to the low-activity state, settling instead onto an elevated equilibrium. A comparable transition has been observed in isolated neocortical preparations when glutamatergic drive is increased experimentally. Partial blockade of glutamate uptake produces network-wide oscillations at approximately $0.015$ Hz, whereas stronger blockade produces only one or two oscillatory events before the network enters a long-lasting NMDA-mediated depolarized state \cite{demarque2004glutamate}. Overall, the experimental preparations and our model converge to predict a range of excitatory drive that supports slow cycling: insufficient excitation cannot sustain the rhythm, and excessive excitation can prolong the active state to the point that alternation is lost. Our delay-free model also predicts that the behavior of infra-slow oscillations is relatively robust to changes in BLA inhibition, with the oscillatory region persisting until relatively weak inhibitory levels. Indeed, BLA slice preparations show that local GABAergic circuitry participates in organizing spontaneous periodic events below $0.1$ Hz \cite{bratsch2024synaptic}, supporting the idea that inhibitory control is in general physiologically relevant in shaping these slow dynamics.

In our model, temporal integration triggers overall changes in cortico-limbic self-interaction and dynamics that transcend the choice of kernel. These changes are broadly consistent with reports from empirical studies of stress-cortisol mechanisms. Cortisol and stress can modify stress-sensitive cortico-limbic parameters: rapid glucocorticoid signaling can suppress GABAergic inhibition of BLA principal cells, effectively shifting local amygdala circuitry toward disinhibition~\cite{di2016acute}, whereas longer-lasting stress and glucocorticoid exposure can induce dendritic retraction and synaptic loss in prefrontal pyramidal neurons, potentially weakening recurrent cortical excitation~\cite{radley2006repeated,wellman2001dendritic,anderson2016prolonged}. 

In turn, greater cortisol responsiveness in humans has been associated with more strongly organized infra-slow dynamics in prefrontal-limbic circuitry~\cite{mujica2014fine}. Chronic hypercortisolism in Cushing’s disease is associated with altered spontaneous low-frequency brain activity, including in prefrontal cortex, with partial normalization after resolution of hypercortisolism \cite{Jiang2017altered}. In rats, a study of CRF-induced stress found that it reorganized oscillatory activity across the infralimbic prefrontal cortex, basolateral amygdala, and hippocampus, producing a delayed reduction of slow-wave activity together with increased theta-, beta-, and gamma-band oscillations \cite{merino2021effects}.  Such studies support the view that cortisol history can reorganize the slow dynamics expressed in cortico-limbic circuits via altering their effective excitation-inhibition balance. What is not yet established experimentally is how these effects depend on the underlying circuit current and past states: the same cortisol history need not have the same dynamical consequence in circuits that begin with different levels of recurrent prefrontal excitation or amygdala inhibition. 

This is precisely where a model can be useful for contextualizing these observations, exploring their possible dynamical consequences, and generating specific hypotheses that can be tested and progressively refined in future work. As an additional example: quasi-periodic infra-slow activity has been observed directly in human electrophysiology: Ko and colleagues identified coherent, quasi-periodic fluctuations in cortical activity centered near $0.015$ Hz~\cite{ko2011quasi}. However, the mechanisms generating such activity and its functional significance remain poorly understood. Our model suggests that quasi-periodicity may not necessarily reflect a distinct intrinsic circuit architecture, but may instead arise from the system's temporal integration, particularly when the relevant history is concentrated around a more sharply localized past time (as in the discrete-delay limit).

\subsection{Limitations and future work}

The main limitation of the present model is its phenomenological nature. Our aim was not to construct a quantitatively fitted representation of cortico-limbic physiology, but rather a proof-of-principle framework for asking how the temporal integration of stress memory can alter effective interactions between prefrontal and amygdala populations, and thereby reshape the resulting dynamics. The functional forms used to represent these interactions were therefore chosen to be as simple as possible while reproducing the qualitative mechanisms of interest. Likewise, the parameterization was guided by known relationships, relative strengths, and physiological constraints reported in the empirical literature, but the exact numerical values were not fitted to data and should not be interpreted as direct physiological measurements. Where possible, we considered parameter ranges compatible with these qualitative constraints and searched within those ranges for the dynamical phenomena of interest. Hence the specific numerical boundaries of the oscillatory and non-oscillatory regimes should be interpreted as model-dependent, with the same applying for the transitions and mechanisms identified here as the main predictions of the framework.

A second limitation is that the modeled circuit is intentionally incomplete. We focused on a minimal prefrontal-amygdala system in order to isolate the consequences of stress-sensitive modulation of recurrent cortical excitation and amygdala inhibition. Emotion regulation, however, emerges from a considerably larger network. The hippocampus, thalamic and hypothalamic structures, neuromodulatory systems, and the HPA axis itself can all contribute to the production, interpretation, and regulation of stress responses. In the present formulation, the influence of these omitted systems is absorbed only indirectly into effective coupling terms, tonic inputs, or the delayed modulation imposed on the cortico-limbic circuit. The model should therefore be viewed as an open subsystem embedded within a substantially larger physiological network rather than as a closed representation of stress regulation.

A related limitation is the absence of direct empirical validation of the model's dynamical predictions. At present, there are no data that simultaneously resolve the evolving cortico-limbic population dynamics and the temporally distributed effects of cortisol with sufficient temporal detail to identify the type of history-dependent interaction represented here. This lack of data is in part the motivation for the present modeling approach. One goal of such models is to identify which measurements would be most informative, and to generate hypotheses about how delayed hormonal modulation should alter neural dynamics under different baseline circuit conditions. Future experimental and modeling work could therefore proceed iteratively: models can suggest specific temporal relationships and parameter dependencies to measure, while new physiological and imaging data can in turn constrain the functional forms, parameter ranges, and delay structures used in the models.

More broadly, understanding how the time course of cortisol action interacts with emotion regulation remains an important and incompletely resolved problem. Cortisol production and action occur within a much wider physiological context and are influenced by factors such as sleep, age, broader hormonal state, autonomic regulation, cardiovascular function, and previous stress exposure. These factors may change not only the magnitude of a cortisol response, but also its timing, duration, and interaction with neural state. A natural next step is therefore to couple the present cortico-limbic model to our previously developed model of HPA-axis dynamics~\cite{kaslik2018stability}, allowing cortisol to emerge from an explicit endocrine feedback system rather than being represented only through an imposed history-dependent modulation. Such a coupled framework would make it possible to study the bidirectional interaction between neural regulation and endocrine stress dynamics, and to ask how changes in either subsystem propagate across timescales to influence resilience, recovery, and persistent dysregulation.

\subsection*{Acknowledgements}

The project received support from a joint NSF-UEFISCDI grant: Project No. \#2408407 (R\u{a}dulescu), Project No. ROSUA-2024-0002 (Kaslik). 

\bibliographystyle{plain}
\bibliography{references}

\end{document}